\documentclass[10pt,twocolumn,letterpaper]{article}

\usepackage[pagenumbers]{cvpr} 

\usepackage[dvipsnames]{xcolor}

\usepackage{bm}
\usepackage{algorithmic}
\usepackage{algorithm}
\usepackage{enumitem}
\usepackage{color}
\usepackage[percent]{overpic}
\usepackage{url}
\usepackage{array}

\definecolor{cvprblue}{rgb}{0.21,0.49,0.74}
\usepackage[pagebackref,breaklinks,colorlinks,citecolor=cvprblue]{hyperref}

\def\paperID{7789} 
\def\confName{CVPR}
\def\confYear{2024}

\title{Pre-capture Privacy via Adaptive Single-Pixel Imaging}

\author{Yoko Sogabe
\qquad
Shiori Sugimoto
\qquad
Ayumi Matsumoto
\qquad
Masaki Kitahara\\
NTT Corporation, Japan\\
{\tt\small \{yoko.sogabe, shiori.sugimoto, ayumi.matsumoto, masaki.kitahara\}@ntt.com}
}

\begin{document}
\maketitle
\begin{abstract}
As cameras become ubiquitous in our living environment, invasion of
privacy is becoming a growing concern. A common approach to privacy
preservation is to remove personally identifiable information from a
captured image, but there is a risk of the original image being
leaked. In this paper, we propose a pre-capture privacy-aware imaging
method that captures images from which the details of a pre-specified
anonymized target have been eliminated. The proposed method applies a
single-pixel imaging framework in which we introduce a feedback
mechanism called an aperture pattern generator. The introduced aperture
pattern generator adaptively outputs the next aperture pattern to avoid
sampling the anonymized target by exploiting the data already acquired
as a clue. Furthermore, the anonymized target can be set to any object
without changing hardware. Except for detailed features which have been
removed from the anonymized target, the captured images are of comparable
quality to those captured by a general camera and can be used
for various computer vision applications. In our work, we target faces
and license plates and experimentally show that the proposed method can
capture clear images in which detailed features of the anonymized
target are eliminated to achieve both privacy and utility.
\end{abstract}
\section{Introduction}
\label{sec:intro}

As a result of technological innovations in networking, semiconductors,
computer vision, and more, cameras have become ubiquitous in our living
environment.
The use of such cameras with computer vision is expected to have various
practical applications.
However, the widespread use of cameras raises concerns about privacy and
may be subject to social backlash and legal restrictions.
Thus, to promote the utilization of computer vision, it is necessary
to overcome privacy and utility trade-offs.

\begin{figure}[t]
\begin{center}
\includegraphics[width=0.95\linewidth]{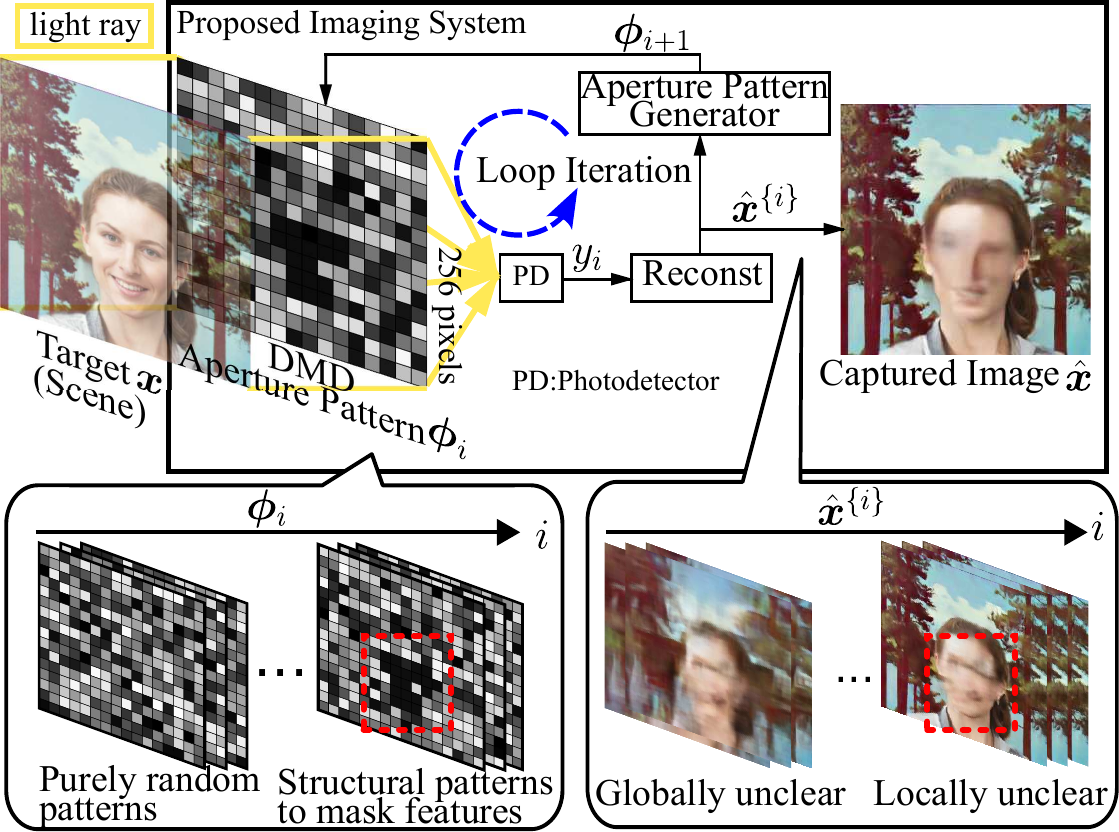}
\end{center}
\vspace{-4mm}
   \caption{Overview of proposed imaging system.
The system captures a single image gradually by repeatedly acquiring
the incident light through the aperture pattern many times based on
 a single-pixel imaging framework.
The entire reconstructed captured image gradually becomes clear.
The next aperture pattern is generated to avoid sampling the
 anonymized target by using the current unclear reconstructed captured
 image.
This feedback mechanism results in the optical elimination of the
 anonymized target.}
\label{fig:fw}
\vspace{-
5mm}
\end{figure}

A common approach to privacy preservation is to remove personal data
from the captured image data after capturing.
However, there is a risk that the data before removal may be leaked.
Such an approach in which personal data is removed after capturing is
called post-capture privacy.
In contrast, pre-capture privacy is an approach based on computational
imaging in which personal data is not captured (either optically or at
the sensor level), which ultimately enhances the level of security.
Prior studies on pre-capture \cite{Zhang2014,pittaluga2016sensor} used
thermal cameras to estimate the location of the face to avoid
sampling it.
These imaging systems were designed by focusing on face anonymization,
and it is difficult to apply them to anonymize anything other than
faces.
With cameras in public places, however, there is a wide variety of
objects that should not be captured, i.e., {\em anonymized targets}.
Examples include faces, textual information (license
plates), fingerprints, and irises.

In this paper, we propose a pre-capture privacy-aware imaging method that
captures images in which the details of the anonymized target are
optically eliminated.
An aperture pattern generator (APG) is introduced in a single-pixel
imaging framework.
The APG implicitly estimates the location of the anonymized target from
the unclear image, which is reconstructed from the already acquired
data, and outputs the next aperture pattern to avoid sampling that
location.
Assuming that the anonymized targets are either faces or license plates,
our quantitative experiments by simulations show that both
privacy and utility can be achieved.
In addition, a prototype imaging
system is assembled to verify its application in the real
world.

Our contributions are as follows:
\begin{description}[wide,itemindent=\labelsep]
\vspace{-2mm}
\item[Pre-capture privacy.]
We introduce imaging that optically excludes detailed features of the
anonymized target by adaptively controlling the aperture.
\vspace{-2mm}
\item[Utility.]
The captured anonymized images are of comparable quality to those
captured by a general camera and can be used for computer vision
tasks.
\vspace{-2mm}
\item[Variability of anonymized targets.]
The anonymized target can be any object other than a face. In this
case, the aperture pattern generation network only needs to be
re-trained using the existing pre-trained recognition model for
 the anonymized target, and there is no need to change the hardware.
\end{description}

\section{Related Work}
Traditionally, the approach to privacy preservation has been
post-capture privacy, but recent advances in computational imaging have
made pre-capture privacy possible.
In relation to our work, we outline methods for pre-capture
privacy and computational imaging techniques that are closely
related to privacy preservation.
\begin{description}[wide,itemindent=\labelsep]
\vspace{-2mm}
\item[Pre-capture privacy for face.] 
The most difficult part of pre-capture privacy is determining
the location of the anonymized target before capturing.
In \cite{Zhang2014} and \cite{pittaluga2016sensor}, a thermal camera
is used to estimate the location of faces.
The thermal camera detects a face silhouette by assuming the
temperature of faces.
Another camera, which can control the shutter pixel-by-pixel, captures
an anonymized image by turning off the shutter at the detected
silhouette.
The captured images are natural, except for the faces, which are masked.
Therefore, they can be used in any computer vision application.
However, such imaging systems have been designed by focusing
on face anonymization,
and it is difficult to apply them to anonymizing anything other than
faces.
\vspace{-1mm}
\item[Pre-capture privacy for specific applications.] 
Some studies have achieved pre-capture privacy by capturing an
image that can only be used for specific applications but the image is
globally degraded to the point that personal data is unrecognizable.
In \cite{Pittaluga2015}, moderately degraded images are captured with a
defocus lens attached to a camera to blur captured images. This preserves
privacy while using the camera for a specific application
such as full-body motion tracking.
In \cite{Hinojosa_2021_ICCV}, a lens's point spread function and human
pose estimation network are jointly trained in an end-to-end fashion.
This makes it possible to degrade private attributes while maintaining
important features for human pose estimation.
In \cite{tasneem2022learning}, an end-to-end trained phase mask is
inserted into the aperture plane to capture an image that is strongly
blurred to protect privacy while enabling depth estimation.
In \cite{bezzam2022learning}, the coded aperture on a lensless camera
and classifier network are jointly trained in an end-to-end fashion.
This makes it difficult for a malicious user to reconstruct the image
while still being suitable for the trained classifier.
Meanwhile,
the captured anonymized image should be usable for not only one task
but various tasks such as people flow analysis,
character recognition, and object detection.
\vspace{-2mm}
\item[Computational imaging in relation to privacy.]
FlatCam~\cite{asif2016flatcam}, the coded aperture
camera~\cite{Llull:13,wang2019privacy}, and single-pixel
imaging~\cite{spi} are based on compressed sensing (CS) theory~\cite{CS}.
CS-based imaging destroys spatial information in the sensor image (raw
image) and visually eliminates privacy in the sensor image.
However, because the original image, which includes personal data, can
potentially be recovered from the sensor image by CS reconstruction
methods, it is not classified as pre-capture.
In \cite{nguyen2019deep}, which is a modified version of FlatCam, facial
information is eliminated in software by detecting the face through CS
reconstruction.  This approach is essentially classified as post-capture
privacy.
\vspace{-1mm}
\end{description}

\begin{figure*}[!t]
\vspace{-2mm}
\begin{center}
\begin{overpic}[width=0.95\linewidth]{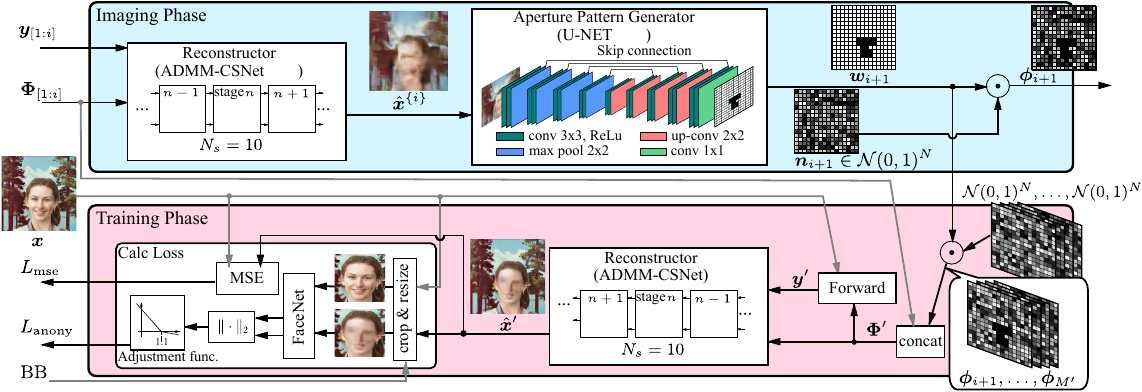}
 \put(23.5,27.6){\scriptsize\cite{8550778}}
 \put(54.0,30.8){\scriptsize\cite{UNET} }
\end{overpic}
\end{center}
\vspace{-6mm}
 \caption{Proposed network architecture.
The aperture pattern generator (APG) implicitly estimates the location
 of the anonymized target from $\hat{\bm{x}}^{\{i\}}$, which is
 reconstructed from the already acquired data, and outputs
 the next aperture pattern $\bm{\phi}_{i+1}$ to avoid sampling that location.
The APG and reconstructor are jointly trained.
The two reconstructors share the network weights.}
\label{fig:net}
\vspace{-3.5mm}
\end{figure*}

To overcome the trade-off between privacy and utility, it is necessary
to be able to set arbitrary anonymized targets and to be able to capture
images without degradation for use in any application.
In contrast to prior studies, the proposed method satisfies all of
these requirements.

\section{Adaptive Single-Pixel Imaging for Privacy}
Figure \ref{fig:fw} shows an overview of the proposed pre-capture
privacy-aware imaging method.
The proposed method is based on a single-pixel imaging (SPI)
framework and introduces a feedback mechanism, called an aperture
pattern generator (APG), using a deep learning model.
SPI gradually captures a single image by repeatedly acquiring incident
light through the aperture pattern, where the entire reconstructed image
gradually becomes clear.
The APG generates the next aperture pattern to avoid sampling
the anonymized target from the current unclear reconstructed image.
This feedback mechanism make it possible to eliminate the anonymized target
optically.
Sec.~\ref{subsec:spi} describes and formulates the principle of
conventional SPI, and Sec.~\ref{subsec:ada} describes how anonymization
is achieved through the feedback mechanism by APG.

\subsection{Single-Pixel Imaging}

\label{subsec:spi}
We explain the mathematical principles, followed by the
optical implementation.
SPI is an imaging method based on CS \cite{CS}.
The target image $\bm{x} \in \mathbb{R}^{N}$ (an
image with a total of $N$ pixels) is not acquired directly but is
-reconstructed from $\bm{y}$ and $\bm{\Phi}$.
First, $\bm{x}$ is optically modulated to a measurement $\bm{y} \in
\mathbb{R}^{M}$ with fewer $M(<N)$ elements using a measurement matrix
$\bm{\Phi} \in \mathbb{R}^{M \times N}$, and then $\bm{y}$ is acquired.
\vspace{-3.0mm}
\begin{eqnarray}
\bm{y}=\bm{\Phi}\bm{x}
\end{eqnarray}
\vspace{-7.0mm} \\
Then $\bm{x}$ is reconstructed from $\bm{y}$ and $\bm{\Phi}$.
\vspace{-2.0mm}
\begin{eqnarray}
\hat{\bm{x}} =\text{Recon}(\bm{y},\bm{\Phi})
\end{eqnarray}
\vspace{-6.0mm} \\
The reconstruction is typically solved by an iterative
algorithm~\cite{tang2009performance}.
A deep unrolled network, which is an algorithm that combines the
advantages of deep learning techniques and traditional iterative
reconstruction algorithms, has also been developed
\cite{zhang2018ista,8550778}.
We choose an unrolled network, ADMM-CSNet~\cite{8550778}, as the
CS reconstructor because of its high computational speed and accuracy.
In this paper, $y_i$ denotes the $i$-th elements of $\bm{y}$,
$\bm{y}_{[1,i]}$ denotes the sub-vector from the 1st to $i$-th elements
of $\bm{y}$,
$\bm{\phi}_i \in \mathbb{R}^{N}$ denotes the $i$-th row vector of $\bm{\Phi}$,
 and $\bm{\Phi}_{[1,i]}$ denotes sub-matrix from the 1st row to the
 $i$-th row of $\bm{\Phi}$.
$M/N$ is referred to as the sampling rate.

SPI involves a photodetector (PD) and digital micromirror device (DMD)
 as shown in Figure \ref{fig:fw}.
The light ray from the target is modulated by the aperture pattern
$\bm{\phi}_i$ displayed on the DMD, and the modulated light is then
acquired in the PD ($y_{i}(=\bm{\phi}_i \cdot \bm{x})$).
The above process is repeated $M$ times to obtain $\bm{y}$.

Additionally, SPI can also be reconstructed using $\bm{y}_{[1,i]}$ at
the $i (< M)$-th acquisition. 
$\hat{\bm{x}}^{\{i\}}(=\text{Recon}(\bm{y}_{[1,i]},\bm{\Phi}_{[1,i]}))$
represents the reconstructed image at the $i$-th acquisition.
When $i$ is small, $\hat{\bm{x}}^{\{i\}}$ is inaccurate, and the
accuracy is expected to increase as $i$ increases.

\vspace{-1mm}
\subsubsection{Block-based CS}
\vspace{-1mm}
\label{subsec:blk}
In the block-based CS \cite{gan2007block}, the target image is
partitioned into small non-overlapping blocks which are acquired
independently but reconstructed jointly.
This can reduce the computational cost of reconstruction.

Suppose that we capture an $L \times L$ image ($N = L \times L$
pixels in total) by dividing it into $B \times B$-pixel blocks ($n =
B \times B$ total pixels in a block).
$N_{b}=N/n$ is the number of blocks.
As $N_{b}$ measurements are acquired every $i$,
$\bm{y} \in \mathbb{R}^{M}$ is acquired for $M'(=M/N_b)$ iterations, and
$\Phi$ is re-defined as an $M' \times N$ matrix.
Additionally, $\bm{y}_i$ denotes the measurements of $N_{b}$ blocks at
the $i$-th acquisition and can be written as
\vspace{-2mm}
\begin{align}
 \bm{y}_i \in \mathbb{R}^{N_{b}} &= \begin{bmatrix}
              y_{i,1} \\ y_{i,2} \\ \vdots \\ y_{i,Nb}
             \end{bmatrix}
    = \begin{bmatrix} \bm{\phi}_{i,1} \cdot \bm{x}_1 \\ \bm{\phi}_{i,2}
       \cdot \bm{x}_2 \\ \vdots \\ \bm{\phi}_{i,Nb} \cdot \bm{x}_{Nb}
       \end{bmatrix}
= \text{Forward}(\bm{\phi}_{i},\bm{x})
\end{align}
\vspace{-6mm}
\\ 
,where $y_{i,j}$ denotes the measurement of the $j$-th block at
$i$-th acquisition, $\bm{\phi}_{i,j} \in \mathbb{R}^{n}$ denotes the $j$-th
block of $\bm{\phi}_i$, and $\bm{x}_{j} \in \mathbb{R}^{n}$ denotes the
$j$-th block of $\bm{x}$.
The measurements from the 1st to $i$-th acquisition are written as
\vspace{-3mm}
\begin{eqnarray}
\bm{y}_{[1:i]} \in \mathbb{R}^{iN_{b}} = [\bm{y}_{1},
 \bm{y}_{2}, \ldots ,  \bm{y}_{i}]^T
\end{eqnarray}
By using $N_{b}$ PDs, each block can be run in parallel so as to reduce
the imaging time to $1/N_b$.

\subsection{Adaptive Aperture Generation for Privacy}
\label{subsec:ada}

\begin{figure}[!t]
\begin{algorithm}[H]
 \caption{Training procedure (\textbf{lines 15--23 are skipped for imaging
 procedure})}
 \label{alg:train}
 \begin{algorithmic}[1]
  \REQUIRE $\bm{x}$, $\text{BB}\text{\footnotesize(Bounding box of the anonymized target)}$
\STATE $\bm{w} \leftarrow \mathbf{1}^{N}$
\STATE $\bm{\phi}_{1} \leftarrow \mathcal{N}(0,1) ^ {N}$
\STATE $\bm{\Phi}_{[1,1]} \leftarrow [\bm{\phi}_{1}]^{T}$
\FOR {$i \leftarrow 1, \ldots ,M'$}
  \STATE $\bm{y}_{i} \leftarrow \text{Forward} (\bm{\phi}_{i},\bm{x})^\dag$
  \STATE $\bm{y}_{[1,i]}  \leftarrow [\bm{y}_{[1,i-1]},\bm{y}_{i}]$
  \IF {$i \in \{ \lfloor K^{n} \rfloor | n \in \mathbb{N}  \}$}
  \STATE $\hat{\bm{x}}^{\{i\}} \leftarrow
  \text{Recon}(\Theta_{R},\bm{y}_{[1,i]},\bm{\Phi}_{[1,i]})$
  \STATE $\bm{w} \leftarrow \text{APG}(\Theta_{G},\hat{\bm{x}}^{\{i\}})$
  \ENDIF 
  \STATE $\bm{w}_{i+1} \leftarrow \bm{w}$
\STATE $\bm{n}_{i+1} \leftarrow \mathcal{N}(0,1) ^ {N}$
\STATE $\bm{\phi}_{i+1} \leftarrow \bm{w}_{i+1} \odot \bm{n}_{i+1}$
  \STATE $\bm{\Phi}_{[1,i+1]} \leftarrow
  [\bm{\Phi}_{[1,i]},[\bm{\phi}_{i+1}]^{T}]$
  \IF {$i \in \{ \lfloor K^{n} \rfloor | n \in \mathbb{N}  \}$}
  \STATE {\scriptsize $\bm{\phi}_{i+2}, \ldots ,\bm{\phi}_{M'} \leftarrow
  \bm{w}_{i+1} \odot (\mathcal{N}(0,1) ^ {N}, \ldots ,  \mathcal{N}(0,1)  ^ {N})$ }
  \STATE $\bm{\Phi}' \leftarrow [\bm{\Phi}_{[1,i+1]},[\bm{\phi}_{i+2},
  \ldots ,\bm{\phi}_{M'}]^{T}]$
  \STATE {\scriptsize $\bm{y}_{i+1}', \ldots , \bm{y}_{M'}' \leftarrow \text{Forward}(\bm{\phi}_{i+1},\bm{x}), \ldots , \text{Forward}(\bm{\phi}_{M'},\bm{x})$}
  \STATE $\bm{y}'  \leftarrow [ \bm{y}_{[1,i]}, \bm{y}_{i+1}', \ldots ,  \bm{y}_{M'}' ]$
  \STATE $\hat{\bm{x}}' \leftarrow \text{Recon}(\Theta_{R},\bm{y}',\Phi')$
  \STATE Calculate $\mathcal{L}_{G}$ using $\bm{x},\hat{\bm{x}}',\text{BB}$,
  and update $\Theta_{G}$
  \STATE Calculate $\mathcal{L}_{R}$ using
  $\bm{x},\hat{\bm{x}}',\hat{\bm{x}}^{\{i\}}$, and update $\Theta_{R}$
  \ENDIF 
\ENDFOR
\RETURN $\hat{\bm{x}} \leftarrow \text{Recon}(\Theta_{R},\bm{y},\bm{\Phi})$
 \end{algorithmic}
\end{algorithm}
\vspace{-5mm}
{\small
$\Theta_{G}$ and $\Theta_{R}$ are the network weights of the APG and
 reconstructor, respectively. \quad
$^\dag$This operation is optical acquisition using the DMD and the PD
 in the real imaging process.}
\vspace{-3mm}
\end{figure}
Introducing a feedback mechanism via an APG into SPI enables anonymization.
In SPI, aperture patterns generated from random normal distributions are
typically used, but in the proposed method, aperture patterns are
derived through a feedback mechanism via the APG.
As shown in Figure \ref{fig:net}, the aperture pattern at $i+1$
($\bm{\phi}_{i+1}$) is adaptively generated from the unclear provisional
reconstructed image at the $i$-th acquisition ($\hat{\bm{x}}^{\{i\}}$)
to avoid sampling the anonymized target.
When $i$ is sufficiently small, $\hat{\bm{x}}^{\{i\}}$ is expected to be
an unclear image, and face silhouettes can be detected even through
individuals cannot be identified.
As a very simple example, it is possible to avoid acquiring detailed
parts of the face by setting $\bm{\phi}_{i+1}$ to zero for the location
of each facial part after the $i$-th acquisition.
Repeating the above process up to the $M'$-th acquisition should
produce a reconstructed image $\hat{\bm{x}}$ in which
only facial features are masked.

Alg.~\ref{alg:train} shows the pseudo-code.
The proposed system captures a single anonymized image by repeatedly
performing the process of the optical acquisition using $\bm{\phi}_{i}$
(line 5) and the adaptive generation of $\bm{\phi}_{i+1}$ by the APG (lines 8--13).
The APG generates $\bm{w}$, which is the sampling weight at
each pixel.
The APG consists of a U-NET deep learning model~\cite{UNET} 
(\#steps=5, \#channels=64) with outputs clipped within the range $[0,1]$
and takes $\hat{\bm{x}}^{\{i\}} \in \mathbb{R}^{N}$ as input and
outputs $\bm{w} \in {[0,1]}^{N}$ as shown in line 9 of
Alg.~\ref{alg:train}.
The next aperture pattern $\bm{\phi}_{i+1}$ is defined as follows:
\vspace{-2mm}
\begin{eqnarray}
\label{eq:phi}
\bm{\phi}_{i+1} = \bm{w}_{i+1} \odot \bm{n}_{i+1}
\end{eqnarray}
\vspace{-5mm} \\
where $\bm{n}_{i+1}$ is a random normal distribution vector (
$\mathcal{N}(0,1) ^ {N}$ ), and $\odot$ denotes an element-wise product.
Because the compressed sensing theory states that a clear image can be
obtained by using $\bm{\Phi}$ of random bases, $\bm{n}_{i+1}$ is used
as the original basis and then is partly attenuated by $\bm{w}_{i+1}$ to
suppress the acquisition of information at each pixel.

In addition, to accelerate the training and imaging process, the
adaptive feedback (line 8--9 of Alg.~\ref{alg:train})
operates only at exponential intervals (line 7),
and the previous $\bm{w}$ is reused (line 11).
Because a random vector $\bm{n}_{i+1}$ is generated at each $i$ (line 12),
a different $\bm{\phi}_{i+1}$ is obtained.
$K$ can be changed in the training and imaging phase.
A too large $K$ causes loss of anonymity.

\subsubsection{Loss Function}
\label{sub:los}
The APG is trained with the following loss function to output
$\bm{w}$ such that only the anonymized target is not sampled.
\vspace{-4mm}
\begin{eqnarray}
\label{eq:loss}
\mathcal{L}_{G} = \alpha L_{\text{mse}} + (1-\alpha) L_{\text{anony}}
\end{eqnarray}
\vspace{-3mm} \\
where $\alpha$ is a balancing parameter.
$L_{\text{mse}}$ and $L_{\text{anony}}$ are used to evaluate image
quality and the degree of anonymity, respectively.
The loss function is evaluated using the target image $\bm{x}$, which
is known at the training phase, and a reconstructed image which depends
on $\bm{w}_{i+1}$.
Note that instead of $\hat{\bm{x}}^{\{i+1\}}$, which is the
reconstructed image at the $(i+1)$-th acquisition, we use $\hat{\bm{x}}'$,
which is the reconstructed image when $\bm{w}_{i+1}$ is reused until
the $M'$-th acquisition.
Since hundreds of acquisitions are required for a
single image, the impact of an aperture pattern ($\bm{w}_{i+1}$) is
small.
To amplify the minute effects of a single $\bm{w}_{i+1}$ and facilitate
learning,
we use the reconstructed image assuming that $\bm{w}_{i+1}$  is
reused until the end ($M'$), namely $\hat{\bm{x}}'$, as shown in lines
16--20 of Alg.~\ref{alg:train}.

$L_{\text{mse}}$ and $L_{\text{anony}}$ are calculated from $\bm{x}$
and  $\hat{\bm{x}}'$ as shown in
the `Training Phase' in Figure~\ref{fig:net}.
$L_{\text{mse}}$ is the mean squared error between $\bm{x}$ and
$\hat{\bm{x}}'$.
$L_{\text{anony}}$ must be small when anonymity is high.
$L_{\text{anony}}$ depends on the anonymized target (face and
license plate), the details of which are defined as
follows:
\begin{description}[wide,itemindent=\labelsep]
\vspace{-1mm}
\item[Face Anonymization.] 
FaceNet~\cite{SchroffKP15} is used as a facial feature extractor.
In FaceNet, the distance of feature vectors from two face images is less
than 1.1 when the two faces are the same individual.
Following this rule, $L_{\text{anony}}$ is calculated as follows:
first, $\bm{x}$,$\hat{\bm{x}}'$, and $\text{BB}$ (bounding box of the
face) are given.
A face image pair is created by cropping $\bm{x}$ and
$\hat{\bm{x}}'$ using $\text{BB}$, and the cropped image pair is
resized to $160 \times 160$ to match the input of FaceNet.
Then we calculate the distance of the feature vectors from the output
of FaceNet (average if there is more than one face) and enter the
distance value into an adjustment function.
The adjustment function is a modified Leaky ReLU
function, i.e.,
\vspace{-1mm}
\begin{equation}
  f(x)=
  \begin{cases}
    - (x-1.1) & \text{if $x<1.1$,} \\
    - 0.01 \times (x-1.1) & otherwise
  \end{cases}
\end{equation}
\vspace{-4mm}
\item[License Plate Anonymization.] 
The basic procedure is the same as that for faces; please refer to
the supplementary material for details.
\end{description}

\subsubsection{Robustness to Reconstruction Attacks}
\label{sub:rob}
\vspace{-1mm}
We need to consider what kind of image attackers will obtain when
all acquisition values, namely $\bm{\Phi}$ and $\bm{y}$, are leaked.
The anonymity level during training is assessed using $\hat{\bm{x}}'$
reconstructed by our training's reconstructor.
However, as attackers may use various reconstruction methods, anonymity
should ideally be robust against any reconstruction method.
To achieve this, the reconstructor is specifically trained for the
$\bm{\Phi}$ property produced by the APG, which should enable it to surpass
the performance of the attackers' reconstructors.
For this purpose, the reconstructor is alternately trained with the APG
using a specific equation, which is line 22 of Alg.~\ref{alg:train}.
\vspace{-3mm}
\begin{eqnarray}
\label{eq:loss2}
\mathcal{L}_{\text{R}} = \frac{1}{N} \| \bm{x} - \hat{\bm{x}}^{\{i\}} \|^{2} +
 \frac{1}{N} \| \bm{x} - \hat{\bm{x}}' \|^{2}
\end{eqnarray}
This perspective is also discussed in the experiment in Sec~\ref{subsec:rob}.

\section{Simulated Experiment}
\label{sec:ex}
\begin{table}[t]
\begin{center}
\small
\begin{tabular}{lwc{1.8cm}wc{1.8cm}}
\toprule
 & Face  & License plate \\
\midrule
Optimizer & \multicolumn{2}{c}{Adam} \\
$l_r$ & \multicolumn{2}{c}{$1.0 \times 10^{-4}$ (halved at every 5 epochs)} \\
\#epochs & \multicolumn{2}{c}{40} \\
\#mini\_batches & \multicolumn{2}{c}{4} \\
Data augmentation & \multicolumn{2}{c}{Random crop, Random resize
     $[0.5:2]$} \\
     &   \multicolumn{2}{c}{Random rotate $[-10^{\circ}:10^{\circ}]$} \\
Training time & \multicolumn{2}{c}{about three weeks} \\
$\alpha$ & $0.999$ & $0.1$ \\
Dataset &\multicolumn{2}{c}{BSDS500~\cite{BSDS500}, DIV2K~\cite{DIV2K}} \\
 & CelebA~\cite{liu2015} & Cars~\cite{krause20133d}\\
\bottomrule
\end{tabular}
\end{center}
\vspace{-5mm}
\caption{Training Parameters}
\label{tbl:trap}
\vspace{-4mm}
\end{table}

We verify the effectiveness of the proposed method through
a simulation experiment in which $\bm{y}_{i} = \text{Forward}
(\bm{\phi}_{i},\bm{x})$ is operated on a computer with the image in the
dataset as $\bm{x}$.
We assume two anonymized targets, a face (Sec.~\ref{sec:face}) and
a license plate (LP) (Sec.~\ref{sec:lp}).
Although it is difficult to compare the proposed method to other
pre-capture privacy-preserving methods, we conduct a quantitative
comparison with the simplest method using defocus blurring with a lens.
To simulate defocus blurring, a $31 \times 31$ Gaussian blur with
$\sigma = 16$ is applied to the input image.
$\sigma$ is adjusted so that the anonymity is almost the same as that of
the proposed method.
We compare `Original', `Defocus', and `Ours', which correspond to
general cameras, cameras with the defocus lens attached, and the
proposed method, respectively.
The target is $256 \times 256$ RGB images ($N=65536$),
and the block size $B$ is 32.
The sampling rate $M/N$ is $0.5$, and $K$ is set to $4$.
Therefore, $M'=512$ and the number of feedback($N_{f}$) is 5.
The programs are written in Python (TensorFlow v2.9.1) and run in Ubuntu
20.04 with an Intel Xeon Platinum 8275CL (memory: 1152GB) and a NVIDIA
Tesla A100 (40GB).
Other training parameters are shown in Table~\ref{tbl:trap}.
All images in the dataset are separated into training and testing images at
a ratio of 9:1.
The anonymized targets (face or LP) are detected from the training
images by the pre-trained detector, and their bounding boxes
($\text{BB}$) are stored in advance.
A total of 64K images are prepared for training.

\subsection{Face Anonymization}
\label{sec:face}

\subsubsection{Training}
The pre-trained
Retinaface\footnote{\url{https://github.com/peteryuX/retinaface-tf2}}~\cite{serengil2021lightface}
is used for the face detector,
the pre-trained
FaceNet\footnote{\url{https://github.com/davidsandberg/facenet}}~\cite{SchroffKP15}
for the facial feature extractor, 
and ADMM-CSNet\cite{8550778} for the CS reconstruction.
The detector and feature extractor are used to prepare the training data
and compute the loss function but are not used in the imaging phase.
ADMM-CSNet is modified to be applicable to block-based CS and
pre-trained using $\bm{\Phi}$ of a random normal distribution matrix in
advance, and the pre-trained weights are used as initial values.
The training images contain a roughly even mix of faces and
non-faces.
The ratio of the number of faces in each image is adjusted to 5:4:1 for
zero, one, and two or more faces.
Since the face images are unaligned, faces appear in various positions.
\vspace{-3mm}
\subsubsection{Results}
\vspace{-1mm}
\begin{figure}[t]
\begin{center}
   \includegraphics[width=0.98\linewidth]{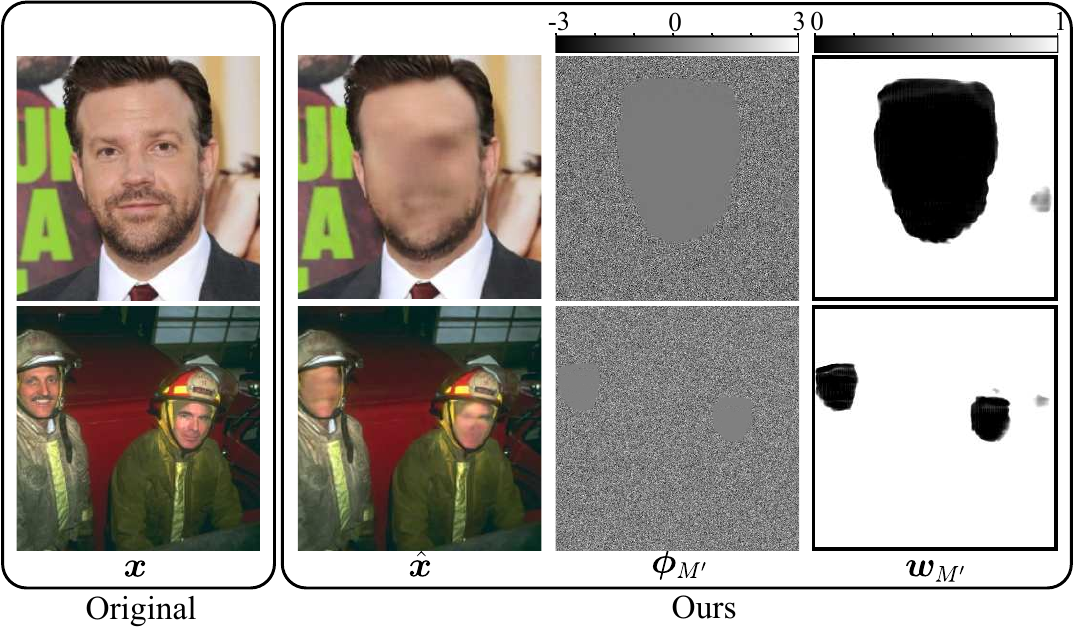}
\end{center}
\vspace{-4.5mm}
 \caption{Images captured from simulations of face anonymization}
\label{fig:face}
\vspace{-2mm}
\end{figure}

\begin{figure}[t]
\vspace{-2mm}
  \centering
 \includegraphics[width=82.0mm]{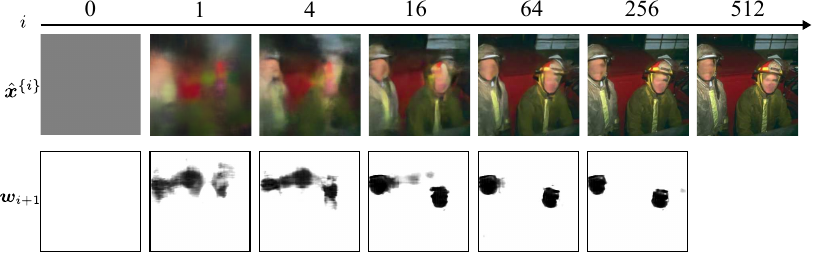}
\vspace{-2mm}
\caption{$\hat{\bm{x}}^{\{i\}}$ and $\bm{w}_{i+1}$ in progress.
 $\hat{\bm{x}}^{\{i\}}$ and $\bm{w}_{i+1}$ are calculated at `$i \in \{
 \lfloor K^{n} \rfloor | n \in \mathbb{N}  \}$' (where $K=4$).
 When $i=0$, $\hat{\bm{x}}^{\{i\}}$ and $\bm{w}_{i+1}$ are the initial
 values (not calculated).
When $i=1$, $\hat{\bm{x}}^{\{i\}}$ is reconstructed, and then the APG
 generates $\bm{w}_{i+1}$ to avoid sampling the faces, although the region may be
 slightly inaccurate.
 The same applies hereinafter at $i=4,16,64,256$.
$\bm{w}_{i+1}$ gradually becomes more accurate.
Finally, when $i=512$, $\hat{\bm{x}} (=\hat{\bm{x}}^{\{M'\}})$
 is reconstructed and outputted as the captured image. }
\label{fig:prog}
\end{figure}

Figure \ref{fig:face} shows the output images.
As shown by $\hat{\bm{x}}$, the clothing, letters, and
background are accurate, while detailed information on the face is
concealed,
making it difficult to identify the person.
Additionally, it remains effective even when multiple faces are presented.
$\bm{w}_{M'},\bm{\phi}_{M'}$ indicates that the face area is set to
zero values to avoid acquiring features.
Figure~\ref{fig:prog} shows the progression of $\hat{\bm{x}}^{\{i\}}$ and
$\bm{w}_{i+1}$ for better understanding of the role of the introduced
APG.
The APG can estimate the location of faces from $\hat{\bm{x}}^{\{i\}}$ and
generate $\bm{w}_{i+1}$ to avoid sampling at the location.
As for computational time, the generation time of $\bm{\Phi}$ per image
is about $0.35$ seconds.

\begin{table}[t]
\begin{center}
\small
\begin{tabular}{@{\extracolsep{4pt}}lcccc@{}}
\toprule
 & \multicolumn{2}{c}{Anonymity}  & \multicolumn{2}{c}{Image quality} \\
 \cline{2-3} \cline{4-5}
Method & LFW($\downarrow$) & \footnotesize{AgeDB-30($\downarrow$)} & PSNR($\uparrow$) &
\hspace{-3.0mm} \begin{tabular}{c}\footnotesize{PASCAL} \\
\footnotesize{VOC2007($\uparrow$)} \end{tabular} \hspace{-3.0mm}
\\
\midrule
Original & 0.999 & 0.987 &  - & 0.6912 \\
Defocus & 0.659 & 0.569 & 21.06 & 0.2535 \\
Ours & 0.675 & 0.558 & 31.64 & 0.6078 \\
\bottomrule
\end{tabular}
\end{center}
\vspace{-4.5mm}
\caption{Results of anonymity and image quality in face anonymization.
`LFW' and `AgeDB-30' indicate AUC value in 1:1 face verification test.
`PASCAL VOC2007' indicates mAP on object detection.}
\label{tbl:face}
\vspace{-3mm}
\end{table}
\begin{figure}[t]
 \vspace{-2mm}
 \begin{minipage}[t]{0.49\linewidth}
  \centering
 \includegraphics[width=45mm]{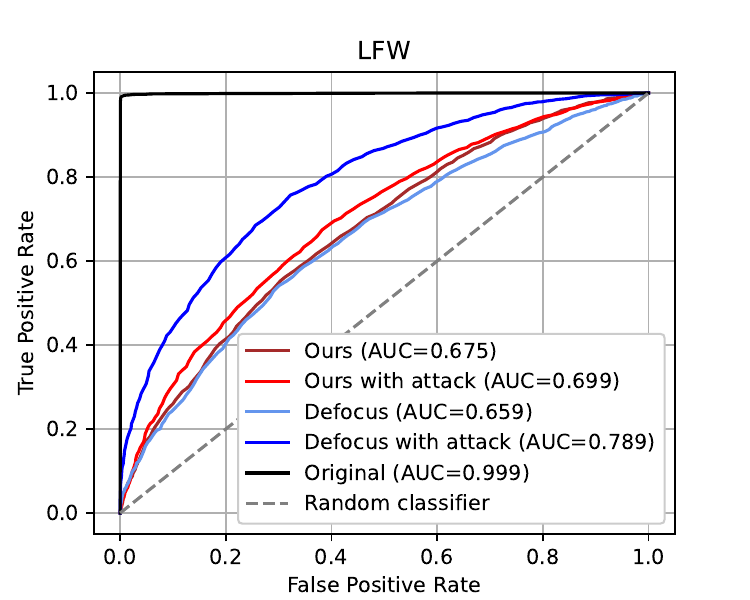}
 \end{minipage}
 \begin{minipage}[t]{0.49\linewidth}
  \centering
  \includegraphics[width=45mm]{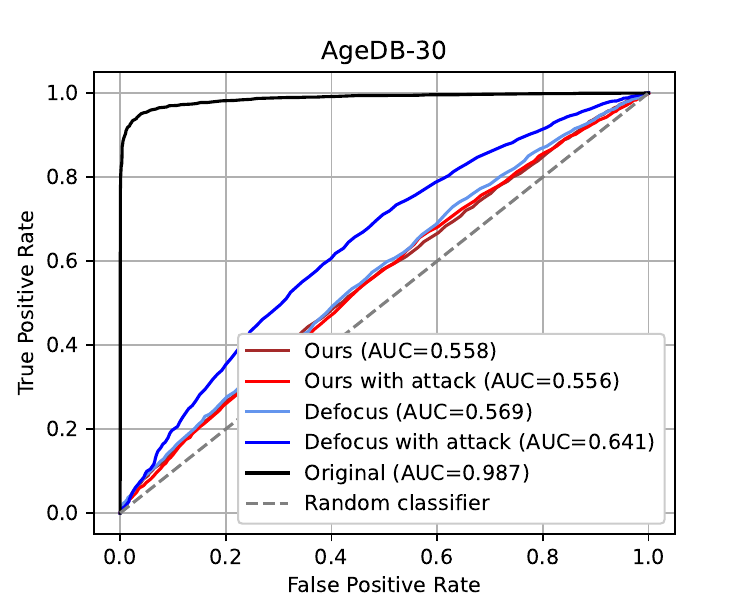}
 \end{minipage}
\vspace{-4mm}
\caption{Anonymity assessment by ROC curve on LFW and AgeDB-30.}
\vspace{-5mm}
\label{fig:roc}
\end{figure}

\begin{figure}[t]
\vspace{-2mm}
 \begin{minipage}[t]{0.49\linewidth}
  \centering
  \includegraphics[width=41.5mm]{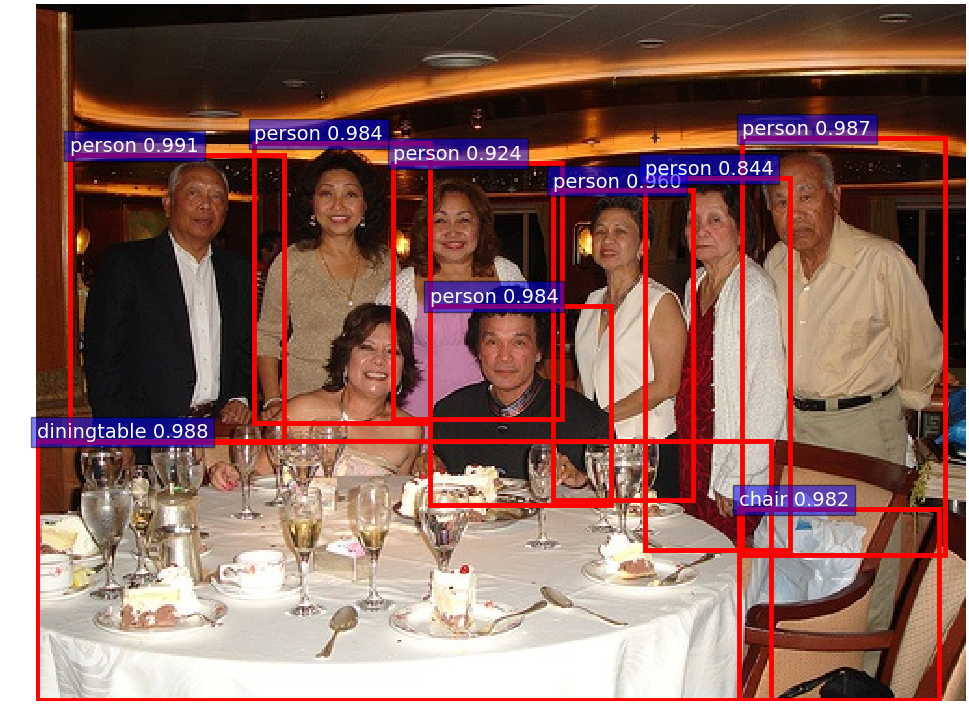}
  \\
  \vspace{-2mm}
 {\footnotesize Original}
 \end{minipage}
 \begin{minipage}[t]{0.49\linewidth}
  \centering
  \includegraphics[width=41.5mm]{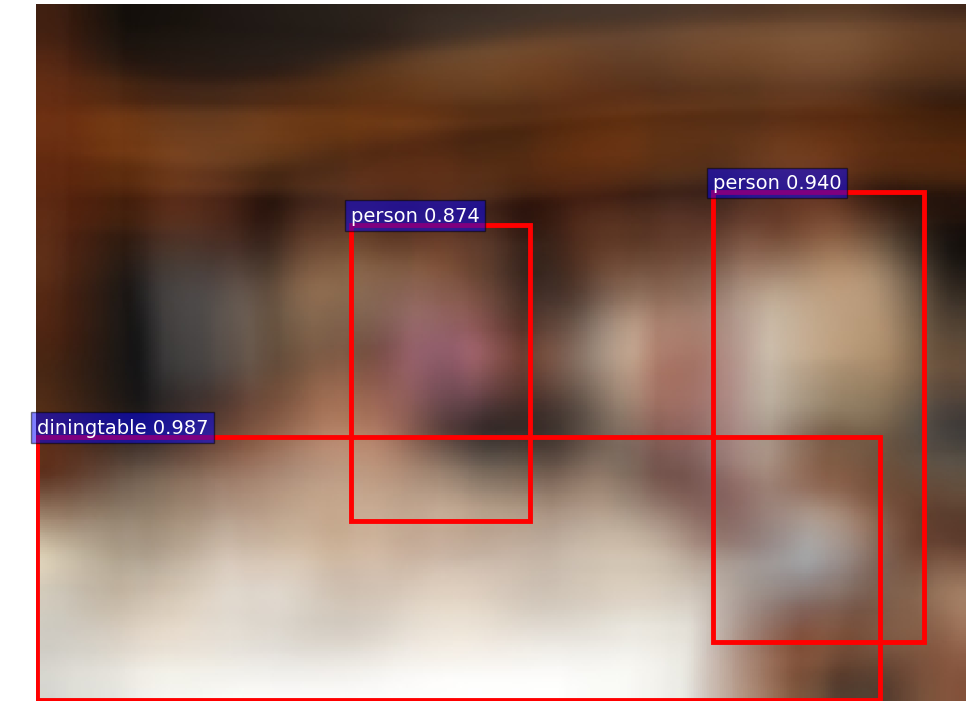}
  \\
  \vspace{-2mm}
 {\footnotesize Defocus}
 \end{minipage}
\vspace{0.5mm}
\\
\centering
 \begin{minipage}[t]{0.49\linewidth}
  \centering
  \includegraphics[width=41.5mm]{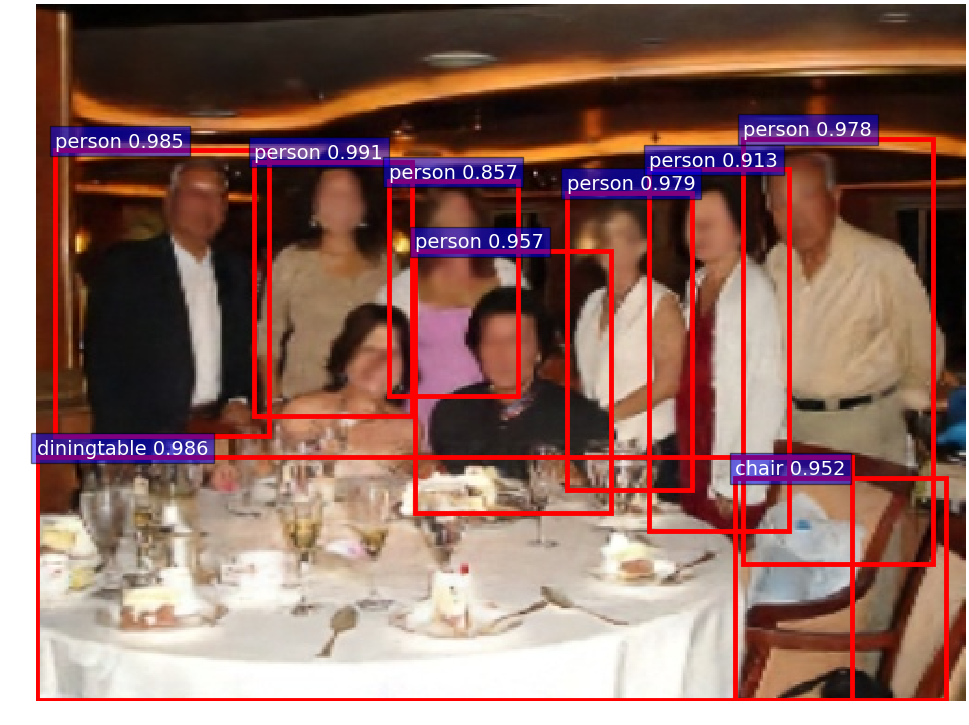}
  \\
  \vspace{-2mm}
 {\footnotesize Ours}
 \end{minipage}
\vspace{-1.5mm}
\caption{Results of object detection.
In `Ours', all objects are detected although the positions of bounding
 boxes are slightly different from that of `Original'.
`Defocus' cannot detect any objects.}
\vspace{-3mm}
\label{fig:det}
\end{figure}

The quantitative evaluation is conducted to assess anonymity and image
quality, the result of which are shown in Table \ref{tbl:face}.
For the anonymity assessment, we perform a face recognition test and
evaluate the area under curve (AUC) of the receiver operating
characteristic (ROC) curve.
We test the LFW~\cite{LFWTech} and AgeDB-30~\cite{AgeDB} dataset using
the pre-trained ArcFace\footnote{\url{https://github.com/peteryuX/arcface-tf2}}
model~\cite{Deng_2019_CVPR}.
A set of `Defocus' and `Ours' images are obtained through Gaussian blur and
simulation of our method, respectively, from the original images in
the dataset.
The images are resized once to $256 \times 256$, each
operation is applied, and then they are resized to back to the original
size.
The column values of `LFW' and `AgeDB-30' indicate the AUC value.
If the faces is recognized completely randomly, the AUC will be 0.5.
Figure \ref{fig:roc} shows the ROC curve.
In AgeDB-30, `Ours' and `Defocus' are close to the random classifier.
The results of the CFP-FP~\cite{cfp-paper} and FGLFW~\cite{FGLFW}
dataset can be found in our supplementary material.
The results show that the proposed method and the defocus lens method achieve a
high degree of anonymity.

The image quality is evaluated by using an image quality metric and
score of object recognition.
For the image quality metric, PSNR is calculated by masking the face
area.
Next, object recognition scores are evaluated to assess whether objects
other than faces are accurately captured.
The dataset PASCAL VOC 2007~\cite{pascal-voc-2007} (20 object classes)
and the detector model
Faster-RCNN\footnote{\url{https://github.com/smallcorgi/Faster-RCNN_TF}}~\cite{ren2015faster}
are used.
The training and testing images of PASCAL VOC 2007 are converted to
`Original', `Defocus', and `Ours', respectively, in advance.
This conversion procedure is the same as that of the face recognition
test.
The Faster-RCNN models is the trained on the training images.
The PASCAL VOC2007 test is performed, and the mean average precision (mAP)
is reported in the `PASCAL VOC2007' column.
The values show that the proposed method clearly captures objects other
than faces.
In contrast,
`Defocus' cannot be used for object recognition due to overall image
degradation.
Figure \ref{fig:det} shows an example of object detection.

\begin{figure}[t]
\vspace{-2mm}
  \centering
 \includegraphics[width=70.0mm]{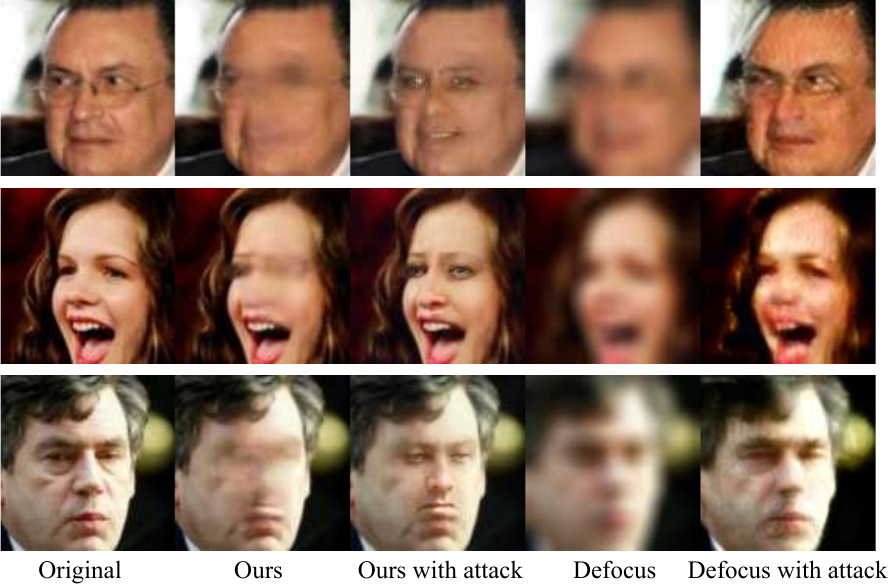}
\vspace{-3mm}
\caption{Image Restoration Attack}
\vspace{-1mm}
\label{fig:res}
\end{figure}

\vspace{-4mm}
\setlist[description]{font=\itshape}
\begin{description}[wide,itemindent=\labelsep]
\item[Image restoration Attacks]
We evaluate the anonymity in the case of an image restoration attack.
Assuming that an attacker can access a set of original ($\bm{x}$) and
reconstructed images ($\hat{\bm{x}}$),
the attacker could train a network to recover the faces.
For this purpose,
we utilize Panini-Net\cite{wang2022panini}, which is the most advanced
GAN-based model for face image restoration and can handle various types
of image degradations.
The training images are converted by using ``Ours'' and ``Defocus'' respectively, and
the model is trained by each of the converted training images.
Figure \ref{fig:res} shows examples of restored images.
With `Ours', Panini-Net frequently restores the face of a noticeably
different person, whereas it is restored quite accurately with `Defocus'.
The quantitative results of the face recognition test under the image restoration
attack are presented as `Ours with attack' and `Defocus with attack' in
Figure \ref{fig:roc}.
The image restoration attack is very effective against `Defocus' but has
little to no effect on `Ours'.
The results show that the proposed method is robust against image
restoration attacks whereas the defocus lens is not.
\end{description}
\vspace{-2mm}
In summary, the results of our quantitative evaluation demonstrate
that only the proposed method achieves anonymity while providing clear
imaging for other objects.

\vspace{-1mm}
\subsection{License Plate Anonymization}
\label{sec:lp}
\vspace{-1mm}

The second anonymized target is set to be vehicle license plates (LPs).
Since the basic experimental procedure follows that of the face version
(Sec.~\ref{sec:face}), we focus on the differences in this section.
The training processes are largely the same; for details, see the
supplementary material.
\vspace{-1mm}
\subsubsection{Results}
\vspace{-1mm}
\begin{figure}[t]
\begin{center}
 \includegraphics[width=0.98\linewidth]{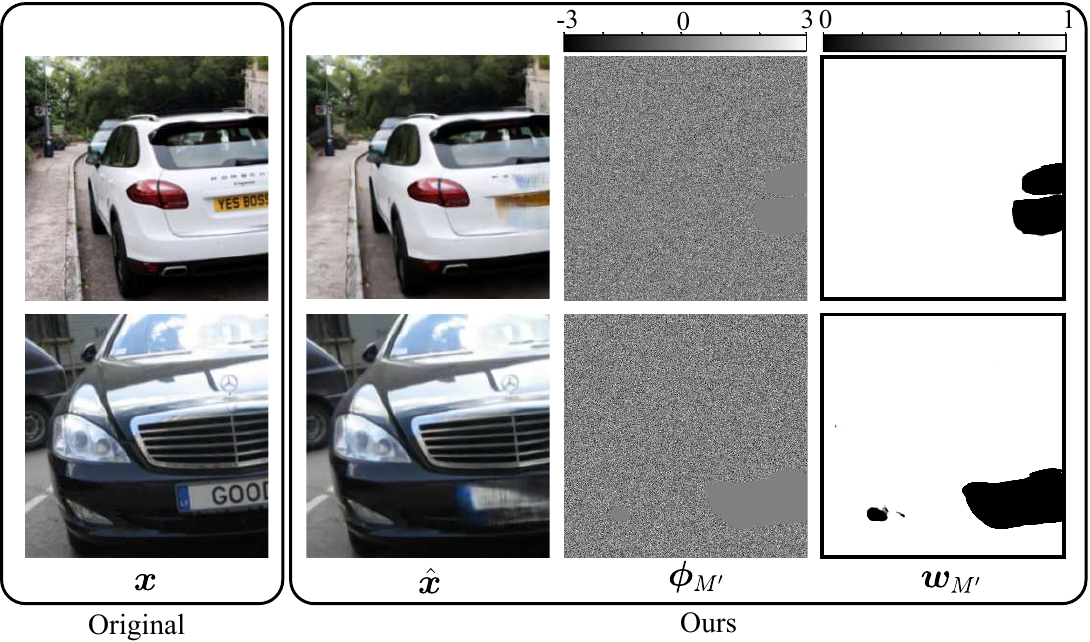}
\end{center}
\vspace{-6mm}
\caption{Images captured from simulations of LP anonymization}
\vspace{-4mm}
\label{fig:car}
\end{figure}

\begin{table}
\begin{center}
\small
\begin{tabular}{@{\extracolsep{4pt}}lccc@{}}
\toprule
 & Anonymity  & \multicolumn{2}{c}{Image quality} \\
\cline{2-2} \cline{3-4}
Method & ALPR($\downarrow$) & PSNR($\uparrow$) & PASCAL VOC2007($\uparrow$)\\
\midrule
Original & 0.715 & - & 0.6912 \\
Defocus & 0.0 & 21.06 & 0.2535 \\
Ours & 0.0 & 31.81 & 0.6117 \\
\bottomrule
\end{tabular}
\end{center}
\vspace{-5mm}
\caption{Results of anonymity and image  quality in license plate
 anonymization.
`ALPR' indicates scores of LP recognition test.
`PASCAL VOC2007' indicates mAP on object detection.}
\label{tbl:car}
\vspace{-2mm}
\end{table}

Figure \ref{fig:car} shows the output images.
As shown by $\hat{\bm{x}}$, detailed information on the
LPs is concealed.
$\bm{\phi}_{M'}$ and $\bm{w}_{M'}$ indicate that the LP area is set to 
zero to avoid acquiring features.

Quantitative evaluation is conducted in term of anonymity and image quality.
Table \ref{tbl:car} shows the results of the quantitative evaluation.
For the anonymity assessment, we follow the
ALPR-Unconstrained\footnote{\scriptsize\url{https://github.com/sergiomsilva/alpr-unconstrained}}
test condition~\cite{silva2018a}, where an LP is considered correct if all
characters are correctly recognized.
ALPR-Unconstrained is used for LP detection and recognition.
As shown in Table \ref{tbl:car}, no LP could be correctly identified in
`Ours' and `Defocus'.
However, the image quality of the proposed method is higher
than that of `Defocus' and is comparable to that of the original.
As in the case of faces, only the proposed method achieves both
anonymity and utility.

\subsection{Reconstruction Attacks}
\label{subsec:rob}

\begin{figure}[t]
\begin{center}
  \begin{minipage}[t]{0.24\linewidth}
    \centering
    \includegraphics[height=18.0mm]{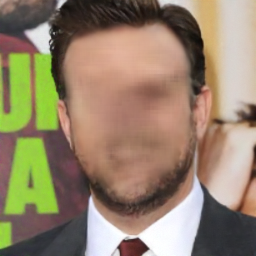}
   {\small ADMM-CSNet \\ (jointly trained)}
  \end{minipage}
  \begin{minipage}[t]{0.24\linewidth}
    \centering
    \includegraphics[height=18.0mm]{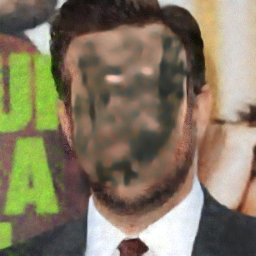}
   {\small ADMM-CSNet \\ {\scriptsize (not jointly trained)}}
  \end{minipage}
  \begin{minipage}[t]{0.24\linewidth}
    \centering
    \includegraphics[height=18.0mm]{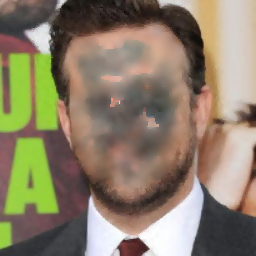}
   {\small ADMM(TV)}
  \end{minipage}
  \begin{minipage}[t]{0.24\linewidth}
    \centering
    \includegraphics[height=18.0mm]{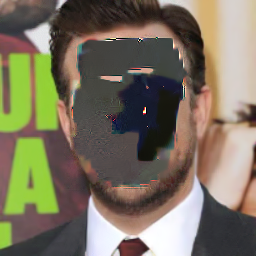}
   {\small PnP(BM3D)}
  \end{minipage}
\end{center}
\vspace{-5mm}
   \caption{Comparison of reconstruction methods}
\label{fig:comp}
\vspace{-5mm}
\end{figure}

In this section, assuming all of the acquired data ($\bm{\Phi}$ and
$\bm{y}$) has been leaked, we compare three reconstruction
methods to verify that the details of the face cannot be recovered.
In Sec.~\ref{sec:face} and \ref{sec:lp}, we evaluated the degree of
anonymity using the reconstructor (ADMM-CSNet) which is trained jointly
with the aperture pattern generator.
As described in Sec.~\ref{sub:rob}, because the reconstructor is trained
to recover data as accurately as possible, including the face, the
evaluation should be reliable.
However, an attacker who obtains $\bm{\Phi}$ and $\bm{y}$ may
reconstruct the target image by any CS reconstruction method.
In this section, we evaluate three different reconstruction
methods: ADMM-CSNet which is not jointly trained, the alternating
directions method of multipliers (ADMM)~\cite{ADMM} with total variation
(TV)~\cite{TV} regularization, and plug-and-play
(PnP)~\cite{venkatakrishnan2013plug} with BM3D~\cite{BM3D}, as shown in
Figure \ref{fig:comp}.
ADMM-CSNet, which is jointly trained, recovers the face most accurately,
indicating that that the anonymity evaluation in Sec~\ref{sec:face}
and \ref{sec:lp} is reliable.

\subsection{Ablation Study}
\label{subsec:abl}
 
\begin{table}
\begin{center}
\small
\begin{tabular}{@{}lcc@{}}
\toprule
Method & Anonymity\\
\midrule
APG & 0.675  \\
segmentation & 0.832\\
\bottomrule
\end{tabular}
\end{center}
\vspace{-3mm}
\caption{Anonymity by segmentation. `Anonymity' indicates the AUC value
 of face recognition on LFW dataset using the pre-trained Arcface model.}
\label{tbl:seg}
\vspace{1mm}
\end{table}

\begin{table}
\begin{center}
\small
\begin{tabular}{@{}lcc@{}}
\toprule
$K$ & Anonymity  & Times[sec] (\#feedbacks)\\
\midrule
1.5 & 0.675 &  0.83 (16) \\
2 & 0.676 &  0.54 (9)  \\
4 & 0.675 &  0.35 (5) \\
8 & 0.683 &  0.26 (3) \\
16 & 0.701 & 0.26 (3) \\
\bottomrule
\end{tabular}
\end{center}
\vspace{-3mm}
\caption{Anonymity by changing $K$.`Anonymity' indicates the AUC value
 of face recognition on LFW dataset using the pre-trained Arcface model.}
\label{tbl:K}
\end{table}

In this section, we discuss the effectiveness of the APG and the parameter
$K$.
As the APG
is similar to a
semantic segmentation network with only two instances (face and
background), we compare the APG with a simple segmentation model to
verify its effectiveness. 
We evaluate the APG trained by Alg.~\ref{alg:train} and the face
segmentation model trained using
CelebAMask-HQ~\cite{CelebAMask-HQ} as shown Table~\ref{tbl:seg}.
The network structure of both models is exactly the same.
The segmentation model compromises anonymity due to its inability to
detect faces in low-quality reconstructed images when $i$ is small.
Additional details can be found in the supplementary material.

If $K$ is too large, the amount of feedback can be reduced but anonymity
would be lost.
Table~\ref{tbl:K} shows how changing $K$ affects anonymity.
When $K$ exceeds 8, the anonymity begins to decline, and speed
does not improve significantly.
However, reducing $K$ to less than 4 does not result in any
anonymity improvement at all.
As a result, we prioritized anonymity and set $K=4$.

\section{Prototype}

\begin{figure}[t]
\begin{center}
\includegraphics[width=0.95\linewidth]{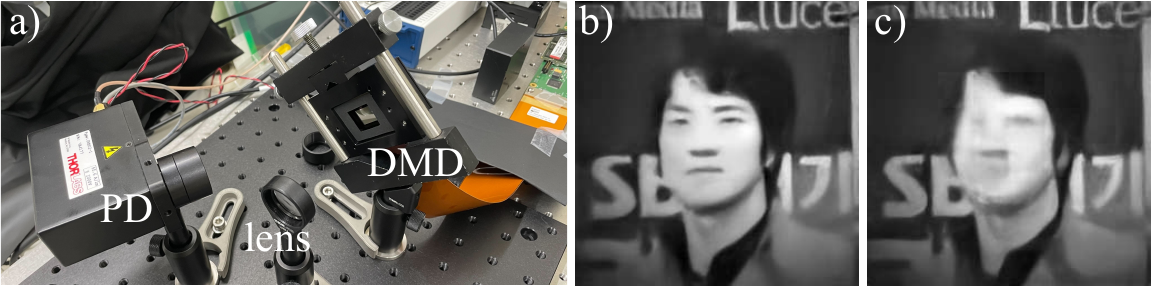}
\end{center}
 \vspace{-4mm}
 \caption{(a) Prototype of the proposed system and images captured by
(b) non-adaptive conventional SPI and (c) our system.
We use a Thorlabs PMM02 as the PD.
The analog to digital converter is a National Instruments USB-6223.
The DMD is a Vialux V7001-VIS for intensity
 modulation.
The objective lens is a Thorlabs LB1901.}
 \vspace{-2mm}
\label{fig:proto}
\end{figure}

As shown in Figure \ref{fig:proto}, we assembled a rough prototype of the
proposed system based on the single-pixel
imaging implementation in \cite{edgar2015simultaneous}.
To simplify implementation, our prototype is degraded in two
aspects compared to the simulated version: the aperture pattern is binary and
the captured images are monochrome.

The subject is a paper printout of one of the images in CelebA, the
sampling rate $M/N$ is $0.75$, and $K=4$.
Since only one PD is used,
the PD sequentially acquires the light from each block by switching off
the blocks other than the target block.
Furthermore, due to inadequate control and synchronization, the system
operates slowly.
As a result, it takes about one minute to capture one image.
Note that the bottleneck is not the processing time of the introduced APG
because the total GPU computing time is less than $0.5$ seconds.
Figure \ref{fig:proto} also shows the captured images in non-adaptive
conventional SPI and in our system,
which demonstrate that the proposed method was effective in the actual
experiment.

\section{Discussion and Conclusion}

We have presented a pre-capture privacy-aware imaging method based on
single-pixel imaging that adaptively generates aperture patterns using a
deep learning model.
The introduced aperture pattern generator outputs
the next aperture pattern by exploiting the data already acquired so as
to exclude features of the anonymized target.
Through simulation experiments on face and license plate anonymization,
we show that our proposed method can anonymize images while maintaining
image quality.

However, the following should be considered with regards to the proposed
method:
\vspace{-2.0mm}
\setlist[description]{font=\normalfont\itshape}
\begin{description}[wide,itemindent=\labelsep]
\item[Real-time imaging.]
Real-time imaging is difficult because thousands of
acquisition values must be sequentially performed for a single image.
The fundamental bottleneck of SPI lies in the operating frequency of the
DMD.
Recent studies~\cite{kilcullen2022compressed,huang202225} have achieved
more than 100 fps by mechanically moving a DMD or modulating light
with LEDs instead of a DMD.
By combining these implementations with the proposed method, real-time
imaging should be feasible.
\vspace{-2.0mm}
\item[Anonymity for reconstruction using temporal adjacency.] 
All experiments in this paper are evaluated assuming that a single image
is recovered from a single $\bm{\Phi}$ and $\bm{y}$.
However, when the proposed method is applied to video, a reconstruction
attack may exploit even multiply pairs of $\bm{\Phi}$ and $\bm{y}$
derived from the previous and next frames.
We have not evaluated the anonymity for such a situation.
\end{description}
\vspace{-1.0mm}
For future work, we plan to improve the hardware implementation for
real-time imaging.
In addition, we plan to conduct further evaluations to expand the scope
of anonymized targets and examine the case where multiple types of
anonymized targets are specified simultaneously.

{
    \small
    \bibliographystyle{ieeenat_fullname}
    \bibliography{main}
}



\onecolumn

{
        \centering
        \Large
        \textbf{\thetitle}\\
        \vspace{0.5em}--- Supplementary Material --- \\
        \vspace{1.0em}
   }


\setcounter{section}{0}
\renewcommand{\thesection}{\Alph{section}}


\section{License Plate Anonymization}

\begin{description}[wide,itemindent=\labelsep]
\item[Losses.]
We define $L_{\text{anony}}$ for license plate (LP) anonymization.
As detailed in Sec. 3.2.1 in our main paper, $L_{\text{anony}}$ must
be small when LPs are concealed.
Unlike face anonymization, which uses the distance between feature
vectors, for LP anonymization, $L_{\text{anony}}$ uses the
mean-squared-error (MSE) within the LP region for simplification.
Specifically, $L_{\text{anony}}$ is calculated as follows.
$\bm{x}$,$\hat{\bm{x}}'$, and $\text{BB}$ (bounding box of
LP) are given.
An LP image pair(s) is created by cropping $\bm{x}$ and
$\hat{\bm{x}}'$ using $\text{BB}$.
Then we calculate the MSE between the cropped LP image pair(s) and
multiply that value by -1.
\item[Training.]
We use the BSDS500~\cite{BSDS500} and DIV2K~\cite{DIV2K} datasets for
general images, as well as the Cars dataset~\cite{krause20133d}, which
includes car images with LPs, as shown in Table 1 in our main paper.
Additionally, we use the ALPR-Unconstrained pre-trained
model~\cite{silva2018a} for license plate (LP) detection and
recognition.
The LPs are detected from the train images by the LP detector, and
their bounding boxes ($\text{BB}$) are stored in advance.
Furthermore, the ratio of car images to other images is adjusted to 1:1.
\end{description}
\vspace{-2mm}

\section{Additional Results of Face Anonymization}
In this section, we report the additional results of the simulated
experiment in Sec.4.1.
\begin{figure}[H]
\vspace{4mm}
\centering
 \includegraphics[width=0.85\linewidth]{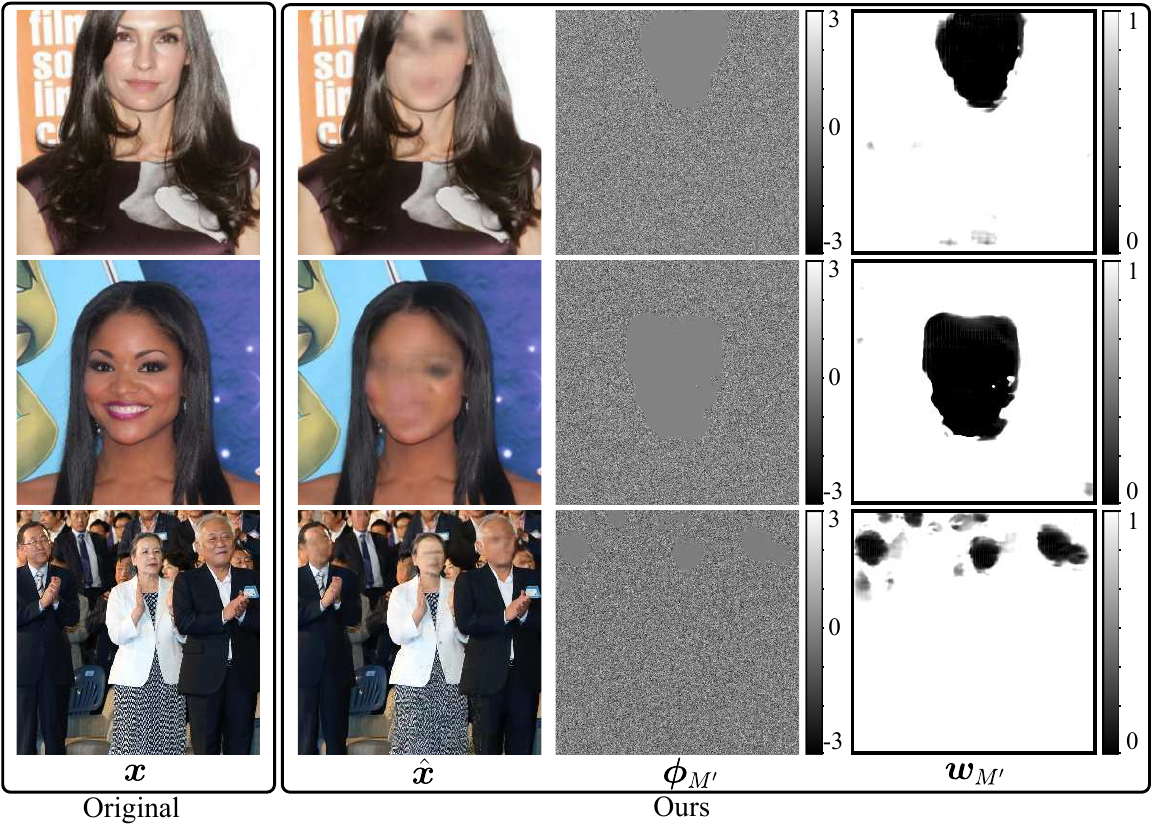}
\vspace{2mm}
\caption{Images captured in simulation of face anonymization}
\label{fig:cap}
\vspace{2mm}
\end{figure}

Figure~\ref{fig:cap} shows the additional output images of the simulated
experiments.
The results show that the proposed method is uniformly effective across
a diverse range of facial features and in various scenarios, such as
different positions and numbers of occurrences, regardless of individual
characteristics.
In addition, the proposed method can precisely capture all of the
objects in the scene other than faces.

\begin{figure}[H]
\vspace{10mm}
 \begin{minipage}[t]{0.495\linewidth}
  \centering
 \includegraphics[width=0.99\linewidth]{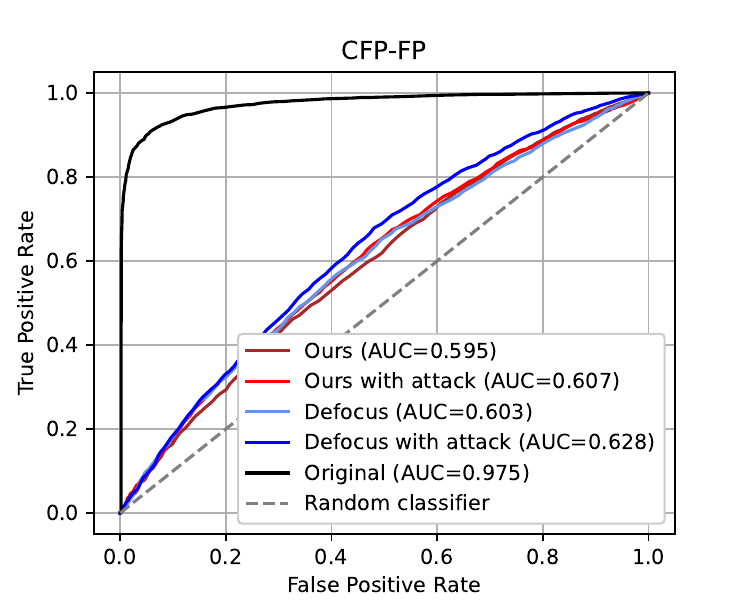}
 \end{minipage}
 \begin{minipage}[t]{0.495\linewidth}
  \centering
  \includegraphics[width=0.99\linewidth]{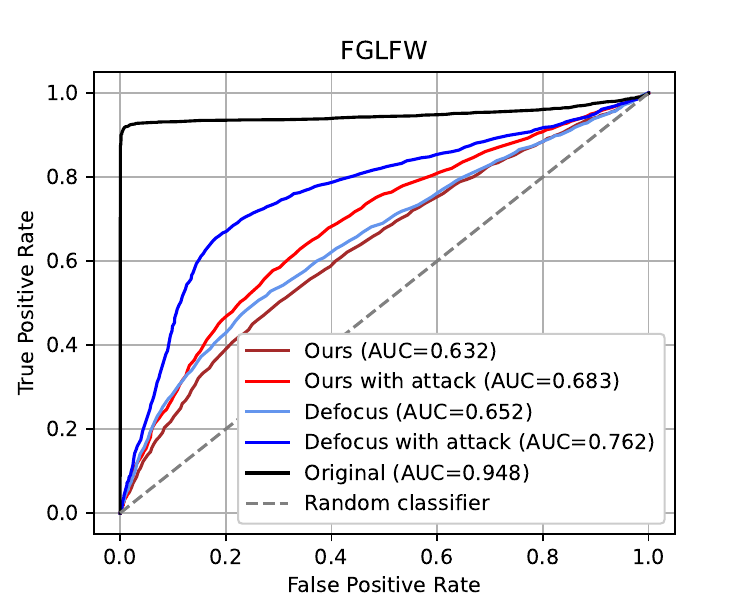}
 \end{minipage}
\vspace{4mm}
\caption{Anonymity assessment by ROC curve on cfp-fp and SLLFW
 dataset. `with attack' indicates values in the case of image
 restoration attacks based on GAN.}
\vspace{3mm}
\label{fig:roc}
\end{figure}

For the anonymity assessment, we perform a face recognition test and
evaluate the area under curve (AUC) of the receiver operating
characteristic (ROC) curve.
In our main paper, we evaluate the Labeled Faces in the Wild (LFW)~\cite{LFWTech} and
AgeDB-30~\cite{AgeDB} datasets for facial recognition (Figure 5).
LFW is the most widely used benchmark dataset with real-world face images,
while AgeDB-30 focuses on age diversity in faces.
In addition, we evaluate the Celebrities in Frontal-Profile (CFP-FP)~\cite{cfp-paper} and
Fine-grained LFW (FGLFW)~\cite{FGLFW} datasets.
CFP-FP focuses on frontal and profile face images of celebrities.
FGLFW, a variation of the LFW dataset, specifically includes
similar-looking face pairs for more challenging face verification tests.
Figure \ref{fig:roc} shows the ROC curves on CFP-FP and FGLFW.
Similar to the results for LFW and AgeDB-30 in our main paper, we verified
that the proposed method can achieve anonymization to a
degree greater than or equal to that of `Defocus' (defocus blurring with
a lens).

\begin{figure}[H]
 \begin{minipage}[t]{0.32\linewidth}
  \centering
  \includegraphics[width=55.0mm]{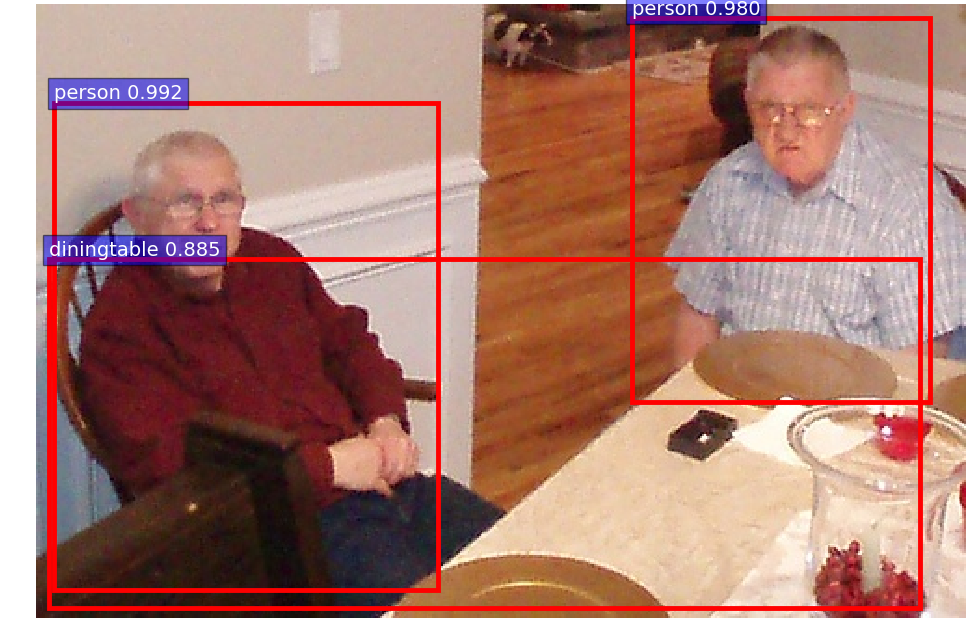}
 \end{minipage}
 \begin{minipage}[t]{0.32\linewidth}
  \centering
  \includegraphics[width=55.0mm]{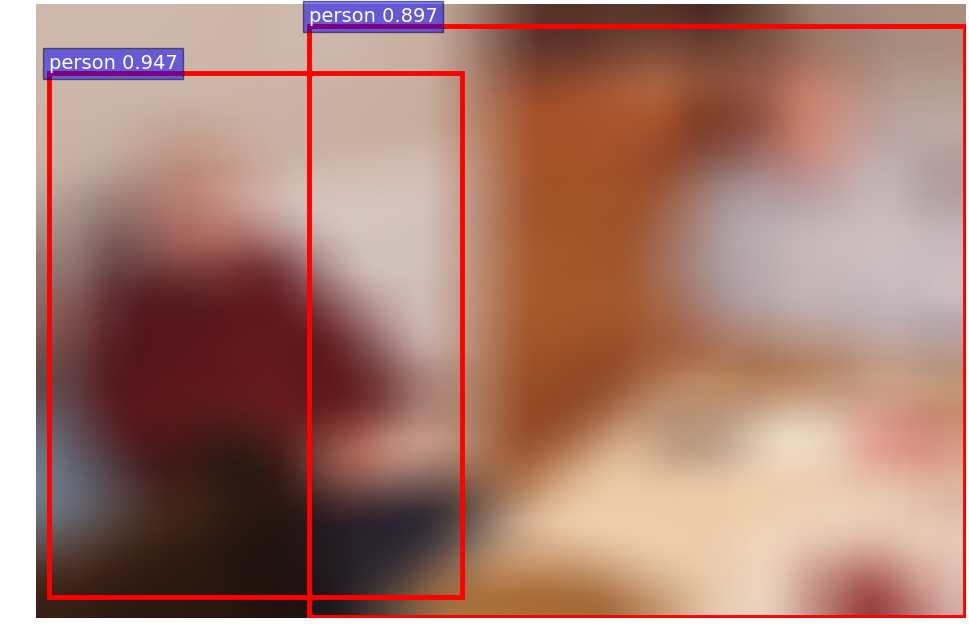}
 \end{minipage}
 \begin{minipage}[t]{0.32\linewidth}
  \centering
  \includegraphics[width=55.0mm]{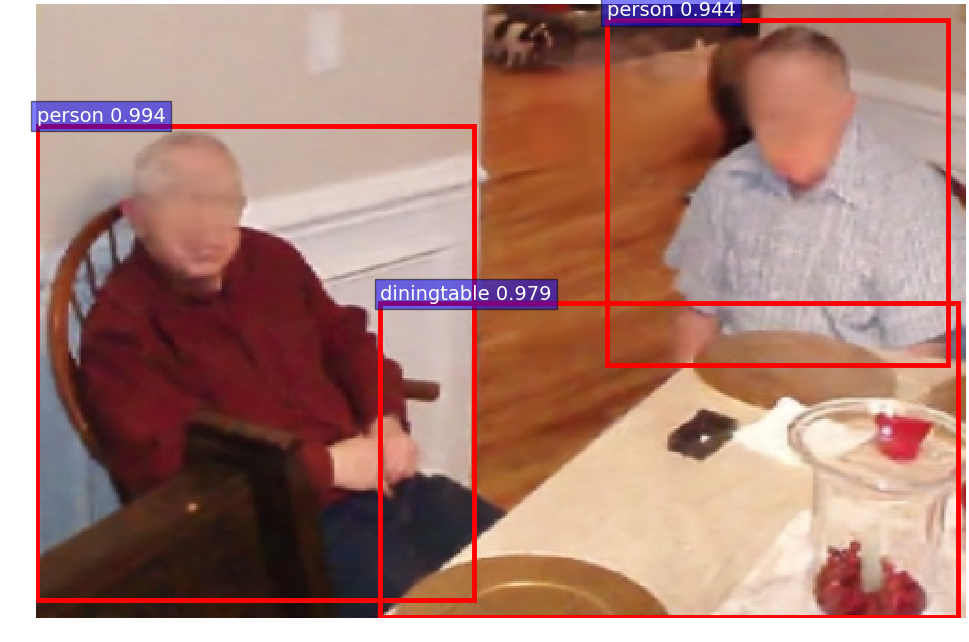}
 \end{minipage}
\\
 \begin{minipage}[t]{0.32\linewidth}
  \centering
  \includegraphics[width=55.0mm]{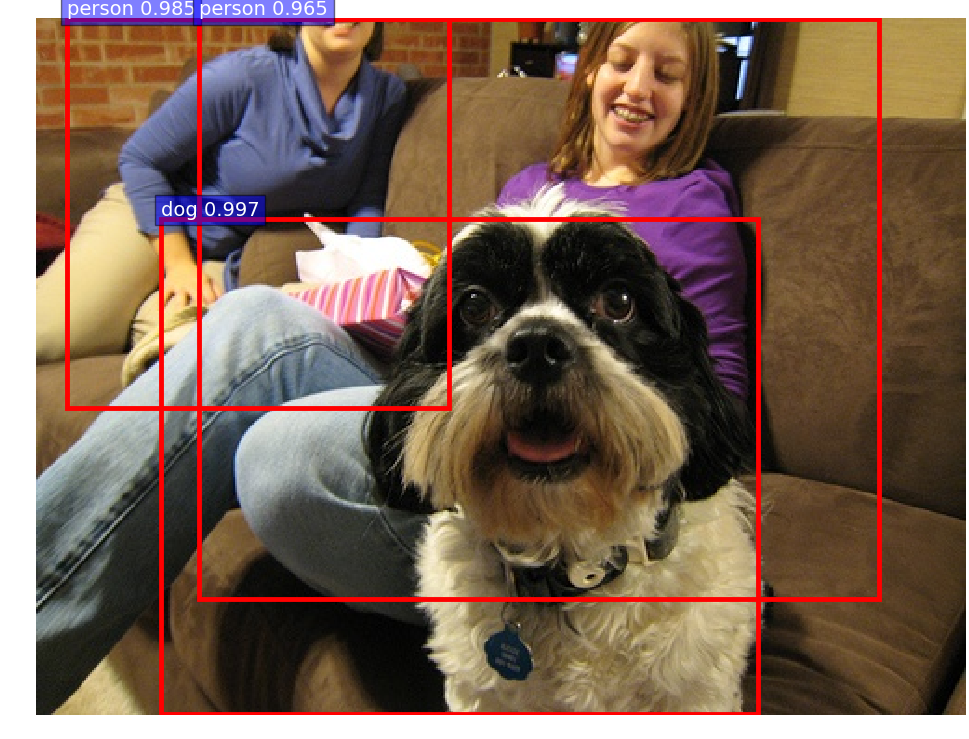}
 \end{minipage}
 \begin{minipage}[t]{0.32\linewidth}
  \centering
  \includegraphics[width=55.0mm]{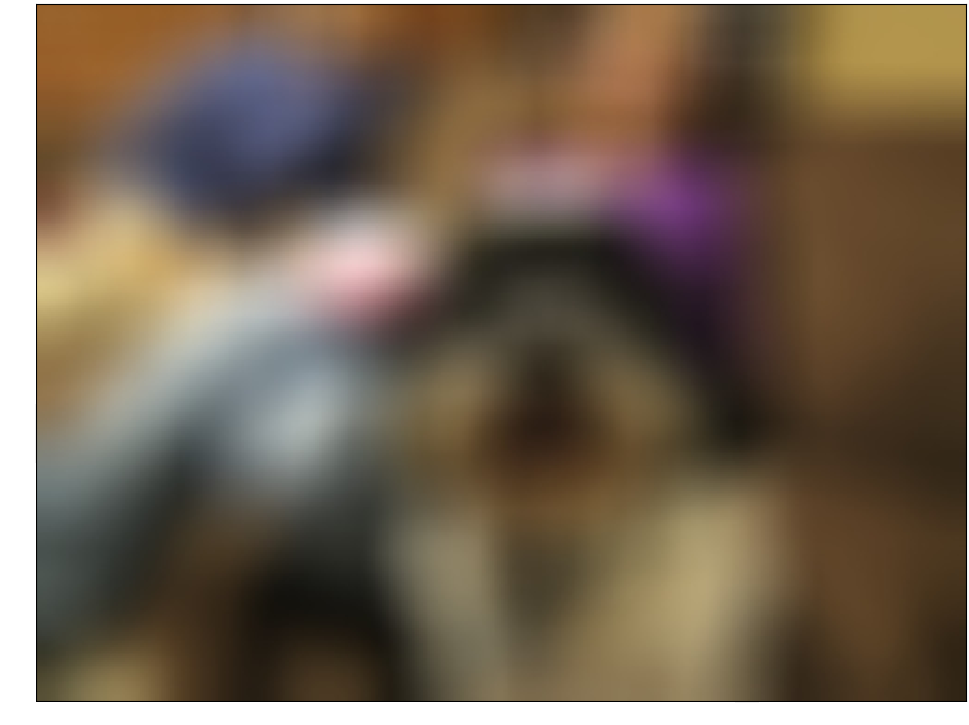}
 \end{minipage}
 \begin{minipage}[t]{0.32\linewidth}
  \centering
  \includegraphics[width=55.0mm]{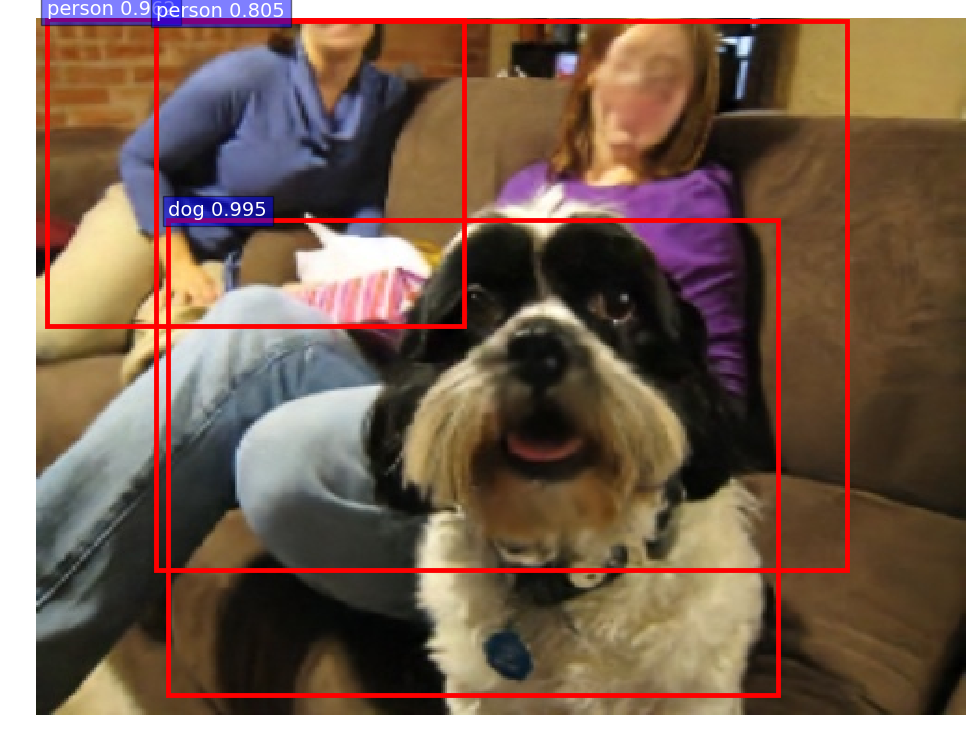}
 \end{minipage}
\\
 \begin{minipage}[t]{0.32\linewidth}
  \centering
  \includegraphics[width=55.0mm]{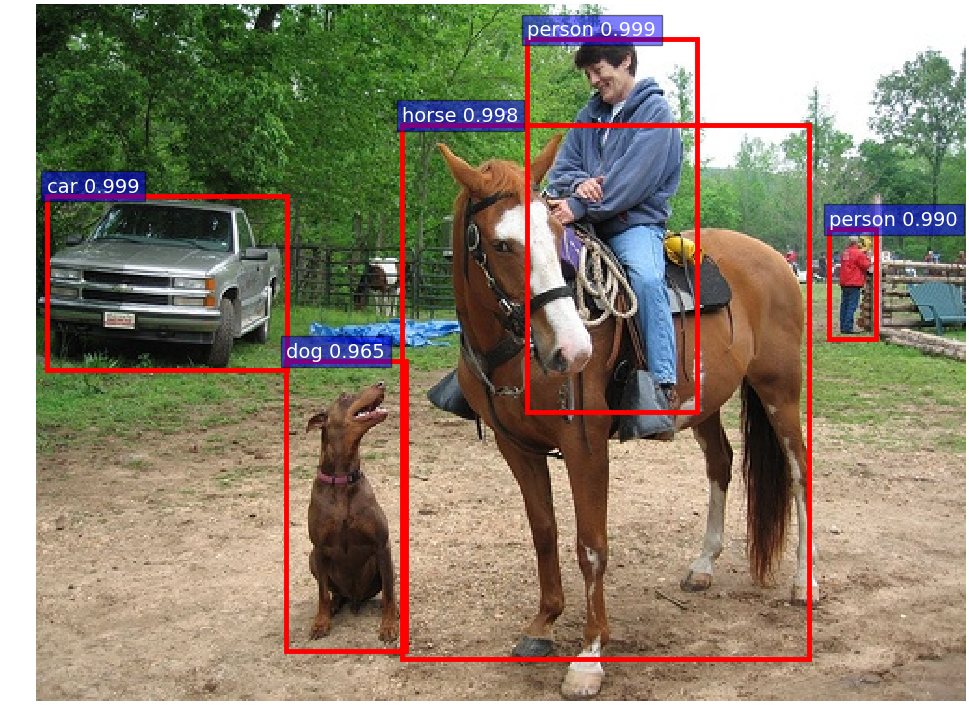}
  \\
  \vspace{-1mm}
 {\small Original}
 \end{minipage}
 \begin{minipage}[t]{0.32\linewidth}
  \centering
  \includegraphics[width=55.0mm]{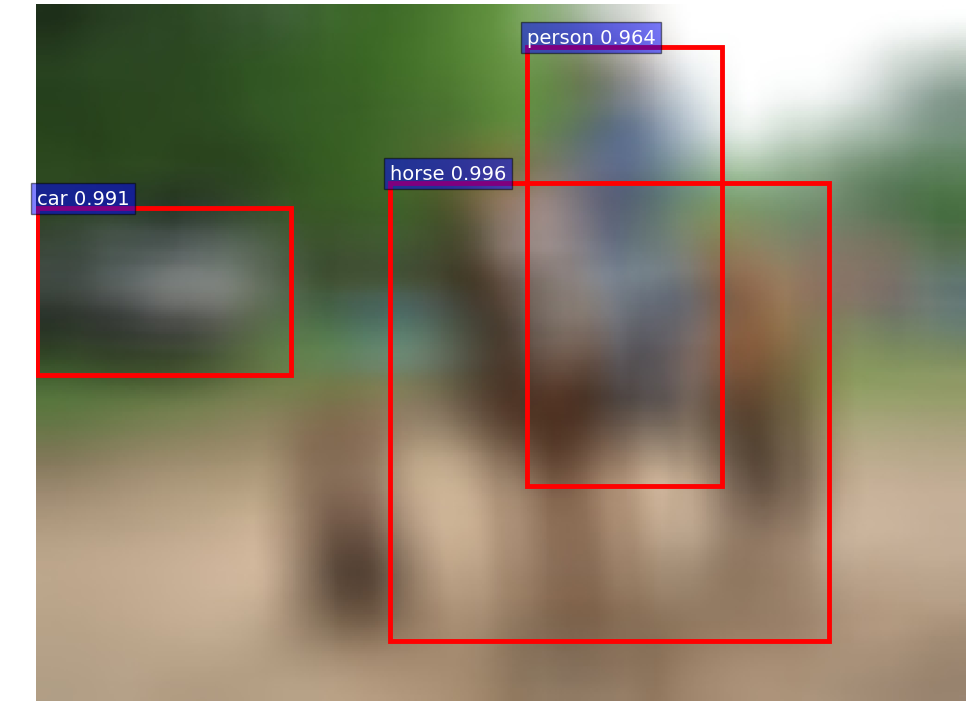}
  \\
  \vspace{-1mm}
 {\small Defocus}
 \end{minipage}
 \begin{minipage}[t]{0.32\linewidth}
  \centering
  \includegraphics[width=55.0mm]{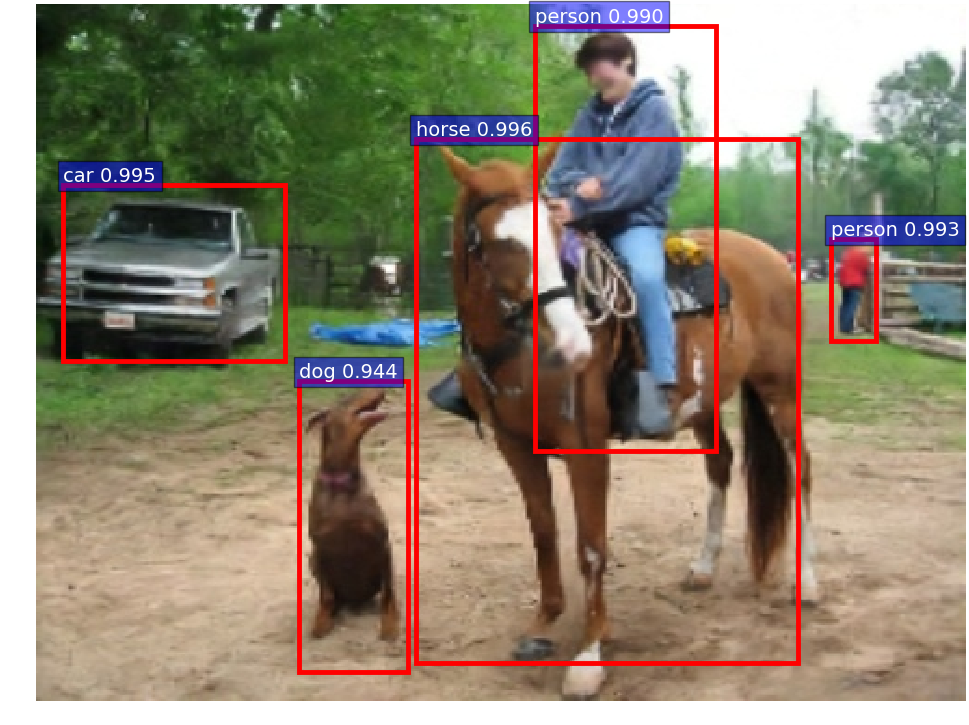}
  \\
  \vspace{-1mm}
 {\small Ours}
 \end{minipage}
\vspace{-1mm}
\caption{Results of object detection.
In `Ours', all objects are detected, although the positions of bounding
 boxes are slightly different from `Original' while the faces are
 concealed.
On the other hand, `Blur' cannot detect any objects.}
\vspace{-3mm}
\label{fig:det}
\end{figure}

\begin{figure}[H]
\vspace{-10mm}
\centering
 \includegraphics[width=0.95\linewidth]{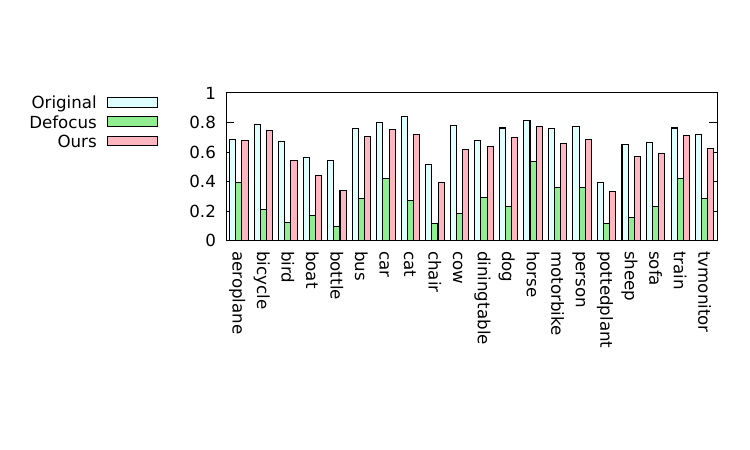}
\vspace{-20mm}
\caption{mAP on each class in PASCAL VOC2007}
\label{fig:cls}
\end{figure}

To assess the accuracy of capturing objects other than faces, we
conducted an object recognition test in our main paper.
Figure~\ref{fig:det} shows additional examples of object detection.
In our approach, denoted as 'Ours', all objects in the scene are
successfully detected.
Moreover, a key aspect of our approach is the ability to maintain
privacy, as we have demonstrated that all faces in the scene are
effectively concealed, providing a balance between object detection
accuracy and privacy protection.
For further analysis, Figure \ref{fig:cls} shows the mAP values on each
class.
In the proposed method, there is no class which is particularly low
quality, and the overall quality is slightly lower than that of
`Original'.
This may be due to the degradation in image quality caused by the
reconstruction in compressed sensing rather than the introduction of the
APG.

\begin{figure}[H]
\vspace{5mm}
\centering
\includegraphics[width=0.85\linewidth]{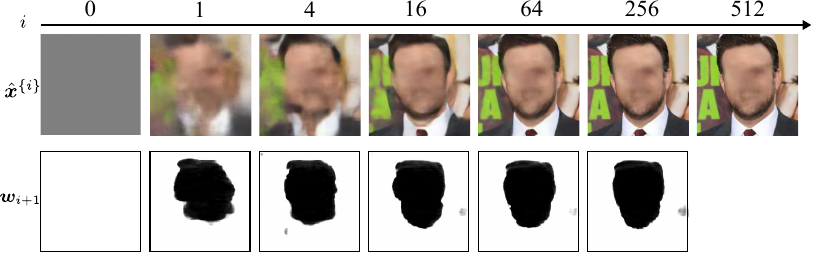}
\\ 
{ APG}
\vspace{6mm}
\\
\includegraphics[width=0.85\linewidth]{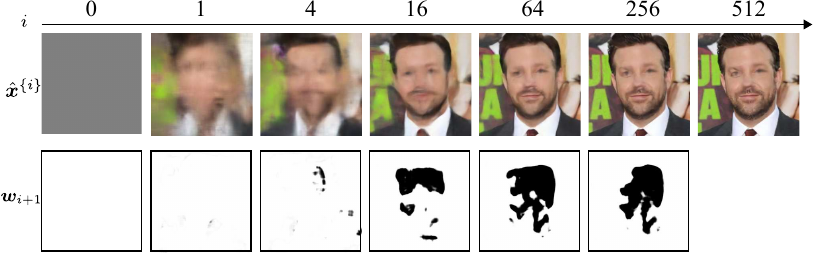}
\\ 
{ Segmentation}
\vspace{2mm}
\caption{Comparison of $\hat{\bm{x}}^{\{i\}}$ and $\bm{w}_{i+1}$ in the APG
 and face segmentation model at each iteration $i$.
$\hat{\bm{x}}^{\{i\}}$ and $\bm{w}_{i+1}$ are calculated at `$i \in \{
 \lfloor K^{n} \rfloor | n \in \mathbb{N}  \}=\{0,1,4,16,64,256\}$'
(where $K=4$).
Initially at $i=0$, both $\hat{\bm{x}}^{\{i\}}$ and $\bm{w}_{i+1}$ are
set to the initial values.
When $i=1$, $\hat{\bm{x}}^{\{i\}}$ is reconstructed, and then
$\bm{w}_{i+1}$ is generated from $\hat{\bm{x}}^{\{i\}}$.
The same applies hereinafter at $i=4,16,64,256$.
Finally, when $i=512$, $\hat{\bm{x}} (=\hat{\bm{x}}^{\{M'\}})$
is reconstructed and outputted as the captured image.}
\label{fig:abl}
\vspace{5mm}
\end{figure}

As described in Sec. 4.4 in our main paper, we evaluated the APG
trained by Alg. 1 and the face segmentation model trained using
CelebAMask-HQ~\cite{CelebAMask-HQ} to assess the effectiveness of the
APG.
While the quantitative results are presented in Table 4 of our main
paper, in this part, we focus on explaining the qualitative
differences between the two by visualizing $\hat{\bm{x}}^{\{i\}}$ and
$\bm{w}_{i+1}$, as shown in Figure~\ref{fig:abl}.
The APG is capable of outputting $\bm{w}_{i+1}$ to avoid sampling the
pixels in the face region by estimating the face's approximate position
from the strongly distorted reconstructed images $\hat{\bm{x}}^{\{i\}}$,
even when $i$ is small.
In contrast, the face segmentation model cannot detect the face in the distorted
reconstructed images when $i$ is small.
Consequently, the face details are unintentionally captured.

\section{Prototype}

As described in our main paper, our current prototype is degraded
in two aspects compared with the simulated experiment: the aperture
pattern is binary and the captured images are monochrome.
To address these, the aperture pattern generator (APG) and
reconstructor are trained with slightly different conditions from
that in Sec. 4.1.
First, the target is set to $256 \times 256$ monochrome images, and all
training images are also converted to monochrome.
Second, the APG and reconstructor are trained with the same conditions
as in Sec. 4.1.
Third, the output of the APG $\bm{w}_{i+1}$ is binarized, and
$\bm{n}_{i+1}$ is changed to random binary values in the imaging phase.

The imaging procedure is as shown in Alg.1 in our main
paper.
Note that, in $\bm{y}_{i} \leftarrow \text{Forward}(\bm{\phi}_{i},\bm{x})$ (line 5), the incident light is
modulated using the DMD, and the modulated light is measured with the PD.
Furthermore, since this is not the training phase, lines 15--23 are
skipped.

In the current implementation, our prototype takes approximately one
minute to capture a single image.
The bottleneck is the acquisition of incident light modulated by 32,768
aperture patterns ($M' \times N_b=512 \times 64$).
The DMD can operate at a maximum frequency of 10kHz, however, 
it becomes a bottleneck as it is significantly slower than the PD and AD
converter.
The expected imaging time based on the DMD's maximum operating frequency
is about 3 seconds, but due to our prototype's inadequate
implementation, the DMD operates at 2kHz, leading to reduced
speed.
Notably, the GPU computation time for introducing the APG is less than
0.5 seconds in total and does not become a bottleneck.
Furthermore, to achieve imaging in less than one second, the
current DMD is insufficient, and the improvements mentioned in Sec.6
of our main paper would be necessary.


\end{document}